\documentclass[fleqn,usenatbib]{mnras}

\usepackage{mathptmx}

\usepackage[T1]{fontenc}
\usepackage{ae,aecompl}
\DeclareRobustCommand{\VAN}[3]{#2}
\let\VANthebibliography\thebibliography
\def\thebibliography{\DeclareRobustCommand{\VAN}[3]{##3}\VANthebibliography}

\usepackage{graphicx}	
\usepackage{amsmath} 
\usepackage{amssymb} 
\usepackage{nccmath}
\usepackage{siunitx}
\usepackage{natbib}
\usepackage{subfigure}
\usepackage[usenames]{color}
\usepackage{multicol}        
\usepackage{pdflscape}	
\usepackage{rotating}
\usepackage{lscape}
\usepackage{threeparttable}
\usepackage[utf8x]{inputenc}
\usepackage{array}
\usepackage{multirow}
\usepackage[normalem]{ulem}
\usepackage{xcolor}

\title[MFA in jets from MYSOs]{Particle acceleration and magnetic field amplification in massive young stellar object jets}

\author[A. T. Araudo et al.]{
Anabella T. Araudo,$^{1,2}$\thanks{E-mail: Anabella.Araudo@eli-beams.eu}
Marco Padovani,$^{3}$
and Alexandre Marcowith$^{4}$
\\
$^{1}$ELI Beamlines, Institute of Physics, Czech Academy of Sciences, 
25241 Doln{\'\i} B\v re\v zany, Czech Republic\\
$^{2}$Astronomical Institute, Czech Academy of Sciences,
Bo\v{c}n\'{\i} II 1401, CZ-141\,00 Prague, Czech Republic\\
$^{3}$INAF-Osservatorio Astrofisico di Arcetri, Largo E. Fermi, 5 - 50125 Firenze, Italy\\
$^{4}$Laboratoire Univers et Particules de Montpellier (LUPM) Universit{\'e} Montpellier, CNRS/IN2P3, CC72, place Eug{\`e}ne Bataillon,\\ 34095, Montpellier Cedex 5, France
}

\date{Accepted XXX. Received YYY; in original form ZZZ}

\pubyear{2015}

\begin{document}
\label{firstpage}
\pagerange{\pageref{firstpage}--\pageref{lastpage}}
\maketitle

\begin{abstract}
Synchrotron radio emission from  non-relativistic jets
powered by massive protostars has been reported, indicating the presence of relativistic electrons and magnetic fields of strength $\sim$0.3$-$5~mG. We study diffusive shock acceleration and
magnetic field amplification in protostellar jets with speeds
between 300 and 1500 km~s$^{-1}$. We show that the magnetic field in the synchrotron emitter can be  amplified by the non-resonant hybrid (Bell) instability excited by the cosmic-ray streaming.  
By combining the synchrotron data with basic theory of Bell instability we estimate the magnetic field in the synchrotron emitter and the maximum energy of protons.
Protons can achieve maximum energies in the range $0.04-0.65$~TeV and emit $\gamma$ rays in their interaction with  matter fields. 
We predict detectable levels of $\gamma$ rays in IRAS~16547-5247 and IRAS~16848-4603.
The $\gamma$ ray flux can be significantly enhanced by the gas mixing due to Rayleigh-Taylor
instability. The detection of this radiation by the Fermi satellite in the GeV domain and the forthcoming Cherenkov Telescope Array at higher energies may open a new window to study the formation of massive stars, as well as diffusive acceleration and magnetic field amplification in shocks with velocities of about 1000~km~s$^{-1}$.
\end{abstract}

\begin{keywords}
acceleration of particles -- radiation mechanisms: non-thermal -- shock waves -- 
stars: jets -- gamma-rays: general
\end{keywords}


\section{Introduction}\label{S:INT}

Stars are formed within dense molecular clouds, accreting matter onto the central protostar with the formation of a circumstellar disc and bipolar jets. These ejections are collimated flows of disc/stellar matter accelerated by magnetic field lines \citep{Blandford_Payne, Shu}, and moving with speeds $v_{\rm j}\sim 300-1500$~km~s$^{-1}$  into the ambient molecular cloud. Molecular matter from the cloud is entrained by the jet, forming molecular outflows. Protostellar jets are thermal radio emitters in most of the cases \citep{Anglada_18Rev}. However, radio emission with negative spectral indices has been detected in several protostellar jets  suggesting a non-thermal nature of the emission
\cite[e.g.][]{Garay_IRAS,Adriana_17,Purser_16,Obonyo_19}. In the particular case of the jet associated with the massive protostar IRAS~18162-2048 and its famous Herbig-Haro (HH) objects HH~80 and HH~81, \cite{Carlos_Sci} reported on polarized radio emission and confirmed that the radiation is produced by the synchrotron process. The detection of synchrotron radiation is an evidence that there is a population of 
mildly-relativistic or relativistic electrons in the jet.

The cosmic-ray (CR) energy density in molecular clouds is poorly known, but it should be at least of the same order of the CR energy density in the interstellar medium, i.e. $\sim 10^{-12}$~erg~cm$^{-3}$ \citep[e.g.][]{Ferriere_01}\footnote{Unless a strong source of CRs contributes to a local enhancement, this number is actually an upper limit as CRs are subject to strong ionization losses while propagating in molecular clouds.}. Therefore, CRs with energy densities larger than $10^{-8}$~erg~cm$^{-3}$ (see Sect.~\ref{S:Rel}) have to be locally accelerated at the jet termination shocks or/and in the internal jet shocks. This result is in agreement with \cite{Marco_15,Marco_16}, who have shown that the large ionization rate estimated from molecular line ratios in L1157-B1 \citep{Podio_14} and OMC2 FIR4 \citep{Ceccarelli_14,Fontani17,Favre+2018} can be due to local acceleration of thermal particles to a relativistic regime in protostellar jets. 

Diffusive Shock Acceleration (DSA) is the most established mechanism to accelerate particles in sources where shock waves are present, from solar flares \cite[e.g.][]{solar_flares} to the outskirts of galaxy clusters \cite[e.g.][]{Petrosian_01}. Magnetic turbulence near the shock allows particles to diffuse back and forth the shock front gaining energy at each crossing cycle \citep{Axford_77, Krymskii_77, Bell_I, Blandford-Ostriker}. The maximum energy that particles can achieve in non-relativistic parallel shocks is usually determined by radiative losses or advection escape from the acceleration region. \cite{IRAS_07} showed that completely ionized and fast protostellar jets can accelerate particles up to TeV energies if Bohm\footnote{The Bohm diffusion regime is obtained when the mean-free path of the particle imposed by angular scattering is equal to its Larmor radius.} diffusion applies. However, in a non-completely ionized medium, damping of Alfv\'en (or magnetohydrodynamic) waves by ion-neutral collisions can reduce the particle maximum energy \citep{Drury_damping, Marco_15, Marco_16}. 

The detection of molecular emission lines such as H$\alpha$, [NII], [SII], and [OIII] in HH objects indicate that protostellar jets are not completely ionized. Emission lines are usually associated with internal (low-velocity) shocks in jets emanating from low-mass protostars. However, in sources like HH~80 and HH~81 molecular lines are co-spatial with synchrotron radiation of locally accelerated electrons \citep{Rodriguez19}. Therefore, strong shocks and scattering waves are present in non-completely ionized plasmas. As a result, short wavelength Alfv\'en waves are unlikely to scatter particles back and forth the shocks due to their damping by ion-neutral collisions. \citet{Tony_04} realized that under certain conditions non-resonant (hereafter NR) hybrid\footnote{The term hybrid comes from the nature of the instability which can be described using both magnetohydrodynamic and kinetic theory.} waves can grow faster than Alfv\'en (resonant) waves. The NR (Bell) instability has two main advantages: $(i)$ the magnetic field is amplified by orders of magnitude and $(ii)$ NR waves are not strongly damped by ion-neutral collisions, even if the maximum growth rate is reduced \citep{Brian_07}.

In this study we show that the magnetic field in the synchrotron emitter of high-mass protostellar jets can be amplified by the streaming of CRs (\citealt{Tony_04}), as in supernova remnants \cite[e.g.][]{Vink-Laming} and jets in Active Galactic Nuclei \citep{4C7426}. 
By assuming that the number of non-thermal electrons and protons is the same, we find that the energy density in non-thermal protons is large enough to drive the Bell instability in the termination region of
protostellar jets. In addition, we show that detectable levels of gamma rays in the GeV domain are expected from protostellar jets.
We consider the sample of non-thermal lobes in the termination region of high-mass protostellar jets studied by \cite{Purser_16} with the Australian Telescope Compact Array (ATCA) facility. We select the sources with radio spectral index $0.3 <\alpha < 0.8$, where the radio flux density  at frequency $\nu$ is $S_{\nu} \propto \nu^{-\alpha}$. 
In addition, \cite{Obonyo_19} found that 6 of the 15 massive young stellar objects (MYSOs) observed with the Jansky Very Large Array (JVLA) show clear evidence of non-thermal emission, with spectral indices $\alpha > 0.42$. From this last sample we select the source IRAS 23262+5834. 
In Table~\ref{Tab_0} we list the MYSOs selected for our study. They are located at a distance $0.7\le d/{\rm kpc}\le 7.8$ from Earth. The jet ionized mass-loss rate is $6\times10^{-8}\le \dot M_{\rm i}/(M_{\odot}$~yr$^{-1})\le 5\times10^{-5}$ and the  kinetic luminosity is $5.3\times10^{33}\le L_{\rm kin}/({\rm erg~s}^{-1}) \le 4.4\times10^{35}$.

\begin{table}
\caption{Non-thermal MYSOs selected from the observational studies carried out by  
\citet{Purser_16} and
\citet{Obonyo_19}. From left to right we list the name of the source in the Green and IRAS catalogues, the distance $d$, and the ionized mass loss rate $\dot M_{\rm i}=X\dot M_{\rm j}$, where $X$ is the ionization fraction 
and $\dot M_{\rm j}$ is the total jet mass loss rate provided in the literature mentioned before. 
The values of $X$ and jet velocity 
fixed by \citet{Purser_16} and
\citet{Obonyo_19}
to compute $\dot M_{\rm j}$ are 
$X=0.2$ and 0.4, and $v_{\rm j}=500$~km~s$^{-1}$. The jet kinetic luminosity computed as $L_{\rm kin}=\dot M_i v_{\rm j}^2/2$ is also listed.}
\begin{tabular}{lccccc}
\hline\hline
Source & IRAS &$d$  &$\dot M_{\rm i}$ & $X$&$L_{\rm kin}$\\
& name &[kpc] & [M$_{\odot}$~yr$^{-1}$] && [erg~s$^{-1}$] \\
 \hline
G263.7434 &  08470-4321 &0.7  & 6.4$\times10^{-8}$ & 0.2&$3.3\times10^{33}$\\
G263.7759 &  08448-4343 &0.7  & 1.3$\times10^{-7}$ & 0.2&$1.1\times10^{34}$\\
G310.1420 &  13484-6100 &5.4  & 1.1$\times10^{-6}$ & 0.2&$9.2\times10^{34}$\\
G313.7654 &  14212-6131 &7.8  & 4.1$\times10^{-6}$ & 0.2&$3.4\times10^{35}$\\
G339.8838 &  16484-4603 &2.7  & 5.3$\times10^{-6}$ & 0.2&$4.4\times10^{35}$\\
G343.1261 &  16547-4247 &2.8  & 3.5$\times10^{-6}$ & 0.2&$2.9\times10^{35}$\\
G114.0835 &  23262+5834 & 4.2 & 4.4$\times10^{-7}$ & 0.4&$3.3\times10^{34}$\\
\hline\hline
\label{Tab_0}
\end{tabular}
\end{table}

The paper is organized as follows: in Sect.~\ref{sec-jet} we describe the physical properties of protostellar jets and the condition for them to be non-thermal emitters. In Sect.~\ref{S:Rel} the magnetic field and the content of relativistic particles is inferred from synchrotron emission. In Sect.~\ref{streaming} we discuss both the resonant and non-resonant instabilities in the context of young stellar jets\footnote{We note that this section is plasma-physics oriented. It is written is a self-contained way. Unfamiliar reader can skip it.}. In Sect.~\ref{sec-MFA} we briefly describe the non-linear saturation of the amplified magnetic field and determine its strength. In  Sect.~\ref{sec-maxE} we compute the maximum energy of protons
and in Sect.~\ref{Sec_gamma} we estimate the $\gamma$-ray flux from our source sample. In Sect.~\ref{sec-conc} we present our conclusions. Gaussian-cgs units are used throughout the paper. 

\section{Protostellar jet physical properties}
\label{sec-jet}

The detection of thermal (free-free) and synchrotron emission from protostellar jets indicates that $(i)$ there are shocks that heat up the plasma and accelerate particles, $(ii)$ the regions in the jet where thermal and non-thermal emission is radiated are at least partially ionized, and $(iii)$ magnetic fields of about 1-10~mG  are needed to explain the detected synchrotron flux if equipartition arguments apply (see Sect.~\ref{S:NTE}). In this section we provide some constraints on the temperature $T_{\rm u}$, the ionization fraction $X_{\rm u}$, the ion density $n_{\rm i}$, and the magnetic field $B_{\rm u}$ upstream of the shock of speed $v_{\rm sh}$
by using the most updated data available in the literature (see Table~\ref{Tab_dens}) and conservation equations. 
We define $X_{\rm u} = (n_{\rm e}+ n_{\rm i})/n_{\rm j}$, where $n_{\rm j}= n_{\rm i} + n_{\rm e} + n_{\rm n}$, $n_{\rm e}$ and $n_{\rm n}$ are the free electrons and neutral densities, respectively, and we assume $n_{\rm e} = n_{\rm i}$ hereafter. 

\subsection{Ionization fraction and jet temperatures}

\citet{Bacciotti_99} estimated the ionization fraction, $X$, in a sample of low-mass protostellar jets. They found that $X$ generally decreases almost monotonically along the jet, hence the effects of the shocked materials over $X$ seems to be minor and we may identify $X$ with $X_{\rm u}$. Typical values of $X_{\rm u}$ between 0.02 and 0.4 have been derived by the authors. \citet{Maurri14} analyzed Hubble Space Telescope arc-second images of the DG Tau jet, hence at a much closer location to the central source than \citet{Bacciotti_99}. They found a low ionization fraction ($\approx 0.02$) at the base of the jet, then increasing to a plateau value of $\approx 0.7$ at a distance of 3\arcsec\ from the central source. The ionization fraction seems then to decrease again. By modeling the optical [OII]/[OI] line ratio in shocks at the base of the HN~Tau jet, \citet{Hartigan04} find $X_{\rm u} \approx 0.7$. \citet{Tesileanu12} developed a magnetohydrodynamic (MHD) model of low-mass protostellar jets and adapted it to the source RW Aur.
They found that $X_{\rm u} \approx 0.1-0.2$ in the central spine correctly fits the surface brightness fluxes in several forbidden lines. They attributed this ionization fraction to irradiation by X-rays from the protostellar magnetic corona. 

The temperature of the unshocked gas is difficult to constrain as line brightness or line ratios are only weakly dependent on it. \citet{Garcia01} investigated low-mass MHD jets launched from accretion discs including ambipolar diffusion heating and several sources of ionization (X-rays and UV radiation as well as collisions). They found a typical jet temperature  $T_{\rm u} \approx 10^4$ K.  

In high-mass protostars both $X_{\rm u}$ and $T_{\rm u}$ are barely known or even constrained. Following \citet{Garcia01} we can reasonably expect the jets from high-mass protostars to have higher mass loss rates and to be denser and cooler than jets from low-mass protostars. Values of $X_{\rm u}$ as small as $10^{-4}$ are found for jets with $T_{\rm u} \approx 10^4$~K. However, other sources of heating like turbulence, MHD wave dissipation, magnetic reconnection, or weak shock heating have not been considered in this work. In a very recent study, \cite{Fedriani_19} found that the ionization fraction in the jet of the high mass protostar G35.20-0.74N is $\sim 0.1$, similar to the values for low-mass protostars.
\cite{Purser_16} assumed $X_{\rm u}= 0.2$ for jets powered by high-mass protostars. For our study we select $0.1\le X_{\rm u}\le 1$ and $T_{\rm u} =10^4$~K in MYSO jets. Hotter (colder) jets likely have higher (lower) ionization fraction.

\subsection{Shocks and densities}
\label{Sec_shocks-dens}

Internal shocks due to a variable ejection velocity is the most popular model to explain the chain of knots in protostellar jets such as HH~111 \citep{Alex_InternalShocks}. In most of the cases, knots are bright H$\alpha$ and [SII] emitters in the optical band indicating shock speeds of few tens of km~s$^{-1}$, much smaller than typical proper motions with speeds of hundreds km~s$^{-1}$. 
Since velocity perturbations are of the order of 10\% of the jet speed, $v_{\rm j}$ \citep{Reipurth_review}, internal shocks are not fast enough to accelerate particles up to very high energies, which is consistent with the fact that most of the knots (and HH objects) are thermal emitters \cite[however, see][for some exceptions]{Adriana_17,Osorio_17}. 

\begin{figure}
\includegraphics[width=0.95\columnwidth]{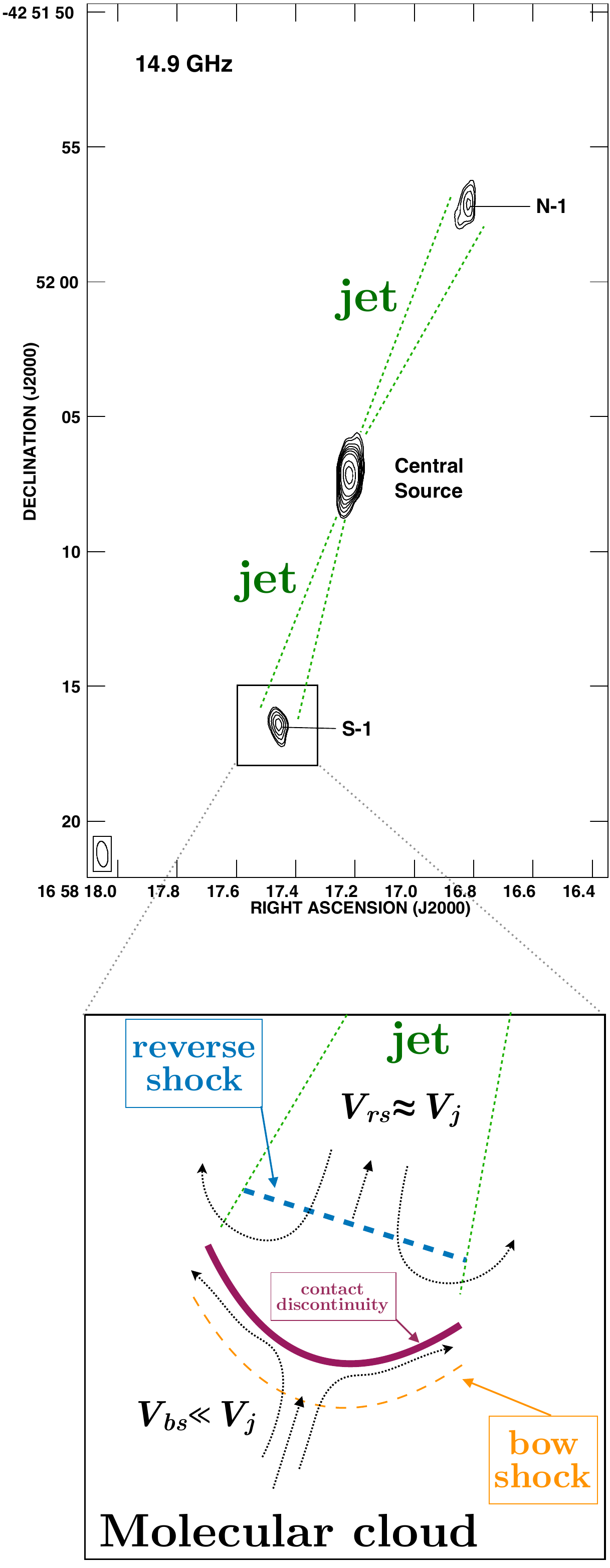}
\caption{Top: The radio image of the source IRAS~16547-4247 (G343.1261) at 14.9~GHz \citep[adapted from][]{LF-05}. Bottom: Sketch of the jet termination region in light protostellar jets. Here, $V_{\rm bs}$, $V_{\rm rs}$, $V_{\rm j}$ refer to the bow shock, the reverse shock, and the jet velocity, respectively.}
\label{sketch}
\end{figure}

Non-thermal emission has been detected in the outer lobes of a handful of protostellar jets, which is more consistent with the presence of a strong adiabatic shock at the location of the non-thermal emitter where synchrotron emitting electrons are 
accelerated. In the jet termination region, where the jet impacts against the external medium (see Fig.~\ref{sketch}), the bow shock moves into the molecular cloud at a speed  $v_{\rm bs} \sim v_{\rm j}/(1 + 1/\beta\sqrt{\chi})$, where $\chi \equiv n_{\rm j}/n_{\rm mc}$ is the jet ($n_{\rm j}$) to molecular cloud ($n_{\rm mc}$) density contrast, and $\beta\equiv R_{\rm h}/R_{\rm j}$, being $R_{\rm j}$ and $R_{\rm h}$  the radius of the jet beam and head, respectively \cite[e.g.][]{Alex_Jorge_WS, Hartigan_MachDisc,Chernin_94}. The reverse shock (or Mach disc) in the jet moves at $v_{\rm rs} = v_{\rm j} - 3 v_{\rm bs}/4$.
In ``heavy'' jets ($\chi > 1$), the bow shock is faster than the Mach disc\footnote{This situation is very similar to the cloudlet model \citep{Norman_Silk} where a fast cloud moves into a molecular cloud \cite[see e.g.][]{Hartigan_MachDisc} or into a slow jet \citep{Yirak_12}.}, whereas in ``light'' jets ($\chi < 1$), the reverse shock is faster than the bow shock. In particular, $v_{\rm rs}\sim v_{\rm j}$ and $v_{\rm bs}\sim v_{\rm j}\sqrt{\chi}$ when $\chi \ll 1$, whereas $v_{\rm bs}\sim v_{\rm j}$ when $\chi\gg1$. As an example, $v_{\rm rs}$ and $v_{\rm bs}$ are $\sim$750 and 250~km~s$^{-1}$, respectively, when $v_{\rm j}=1000$~km~s$^{-1}$ and $\chi = 0.1$. Proper motions of about $300-1400$~km~s$^{-1}$ have been observed in protostellar jets \cite[e.g.][]{Marti_93, Masque_12}, and therefore we can expect jet velocities comparable to or larger  than these values. 

The detection of GeV photons from classical novae \citep{Fermi_Novae} indicates that radiative shocks with velocities $\le 1000$~km~s$^{-1}$ can accelerate particles up to $\sim$10~GeV \citep[see e.g.][]{Vurm_18}. The present study concentrates, however, on adiabatic (i.e. non-radiative) shocks. The condition for a shock with velocity $v_{\rm sh}$ to be adiabatic is $q\equiv l_{\rm th}/R_{\rm j} >1$ \citep{Blondin_90}, where we define $R_{\rm j}$ as the radius of the section of the jet at the position of the hotspot.
The thermal cooling length is
\begin{equation}\label{l_th}
    \frac{l_{\rm th}}{\rm cm} \simeq 6.9 \times10^{16}
    \left(\frac{n_{\rm u}}{10^4\,\rm cm^{-3}}\right)^{-1}\left(\frac{v_{\rm sh}}{1000\,\rm km\,s^{-1}}\right)^\frac{9}{2}
\end{equation}
\cite[e.g.][]{Raga_Xrays}, where $n_{\rm u}$ is the density upstream of the shock. The condition $q>1$ can be rewritten as $v_{\rm sh} > v_{\rm sh, ad}$, where
\begin{equation}\label{cond_th0}
    \frac{v_{\rm sh, ad}}{\rm km~s^{-1}} \simeq 650
    \left(\frac{n_{\rm u}}{10^4\,\rm cm^{-3}}\right)^\frac{2}{9}
    \left(\frac{R_{\rm j}}{10^{16}\,\rm cm}\right)^\frac{2}{9}.
\end{equation}
The reverse- to bow-shock cooling parameter ratio is $q_{\rm rs}/q_{\rm bs} = l_{\rm rs}/l_{\rm bs}\sim \chi^{-3.25}$ indicating that the termination region of light jets ($\chi <1$) is composed by an adiabatic Mach disc and a radiative bow shock.

\begin{figure}
\includegraphics[width=\columnwidth]{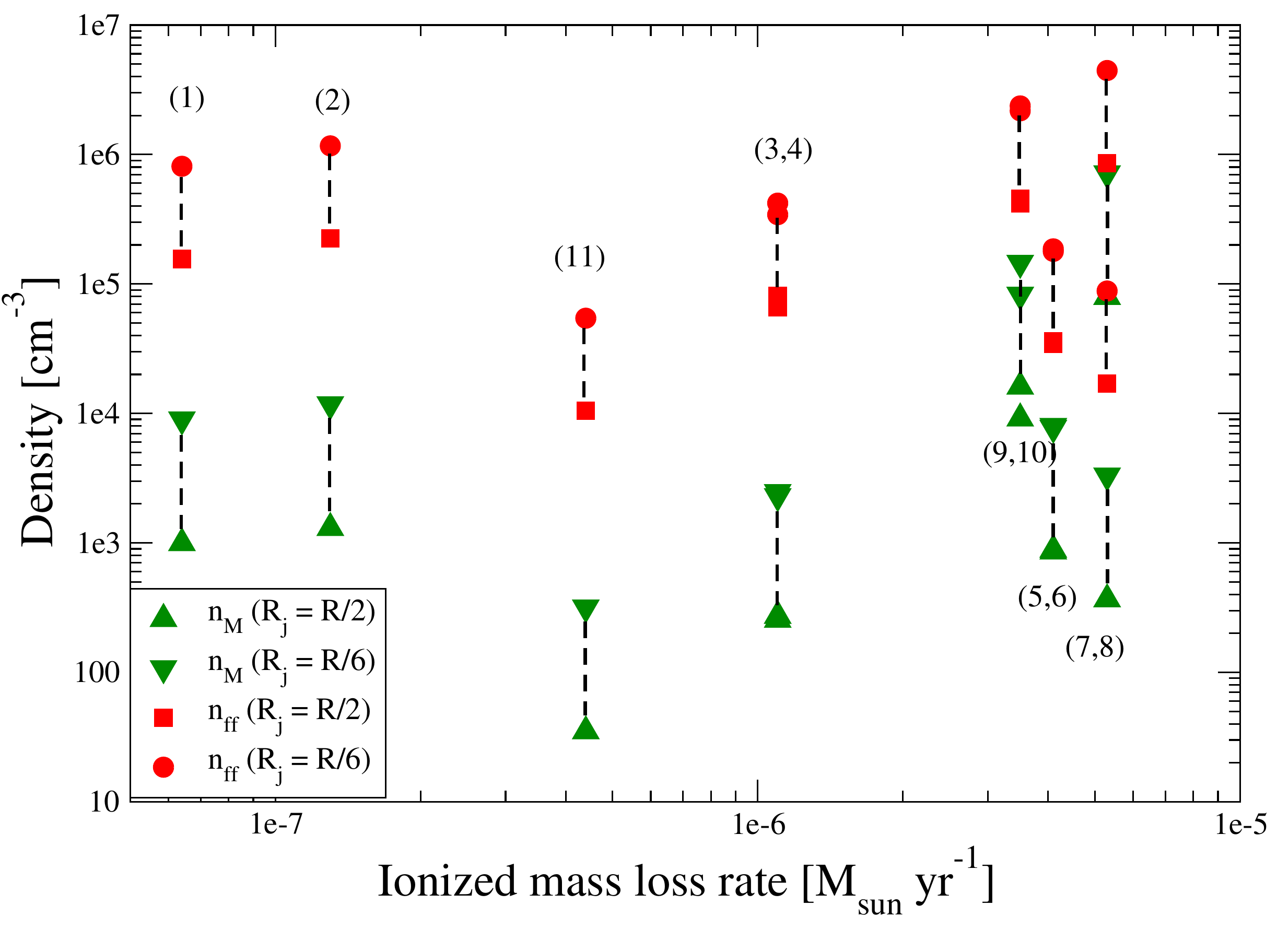}
\caption{Jet ion density, $n_{\dot M}$, for the sources listed in Table~\ref{Tab_dens} as a function of the ionized mass loss rate assuming   $R_{\rm j}=R/2$ (green triangles up) and $R/6$ (green triangles down). Red squares and circles indicate the value of 
$n_{\rm ff}$ when $R_{\rm j}=R/2$ and $R/6$, respectively. 
\label{vshock-min}}
\end{figure}

The jet ion density can be estimated as 
\begin{equation}\label{dens_z}
\frac{n_{\dot M}}{\rm cm^{-3}} \approx 150
\left(\frac{\dot M_{\rm i}}{10^{-6}\,\rm M_{\odot}\,yr^{-1}}\right)
\left(\frac{v_{\rm j}}{1000~\rm km\,s^{-1}}\right)^{-1}
\left(\frac{R_{\rm j}}{10^{16}\,\rm cm}\right)^{-2}
\end{equation}
\cite[e.g.][]{Adriana_17}.
Owing to the large uncertainties in the different parameters we neglect the contribution of helium in Eq.~\eqref{dens_z}. In Fig.~\ref{vshock-min} we plot
$n_{\dot M}$ for $R_{\rm j}=R/2$ and $R/6$, where $R$ is the average linear size of the hotspot also listed in Table~\ref{Tab_dens}. The jet mass loss rate of ionized matter $\dot M_{\rm i} \propto v_{\rm j}$ is assumed to be constant along the jet and then $n_{\dot M}$ turns out to be independent of the jet velocity. By inserting Eq.~(\ref{dens_z}) in Eq.~(\ref{cond_th0}) we find that $v_{\rm sh,ad}\propto R_{\rm j}^{-2/9}$. 
In Table~\ref{Tab_dens} we list 
$n_{\dot M}$ and  $v_{\rm sh,ad}$
assuming $R_{\rm j} = R/2$. 
We note that $n_{\dot M}$ is a rough estimation of the jet ion density given the uncertainties in the values of $\dot M_{\rm i}$ and $R_{\rm j}$. In particular, 
light adiabatic jets form a cocoon and therefore the size of the non-thermal lobe at the jet head is expected to be $R_{\rm h}> R_{\rm j}$. \cite{Krause_03} found that $(R_{\rm j}/R_{\rm h})^2 \sim 0.1-1$  when $\chi>0.01$; this gives  an increment in the jet density by a factor of $\sim 10$ when $R_{\rm j}=R/6$
instead of $R/2$ (see Fig.~\ref{vshock-min}). Nevertheless, the derived jet density is likely lower than the typical density values of $\sim10^5$~cm$^{-3}$ in the molecular clouds where massive stars form \citep{Hennebelle:12} giving $\chi < 1$. Therefore
$q_{\rm rs}>q_{\rm bs}$ and we should rather expect non-thermal sources in light jets ($\chi < 1$), where the reverse shock is faster than the bow shock. 

In order to check if synchrotron emission comes from the reverse-shock rather than from the bow-shock downstream region, we compare
the synchrotron ($\epsilon_{\rm synchr,\nu}$) 
and free-free ($\epsilon_{\rm ff,\nu}$) emissivities. The temperature of the plasma immediately downstream of the shock with
compression ratio $r=4$ is $T_{\rm d} \sim 2.3\times10^7~(v_{\rm sh}/1000\,{\rm km \, s^{-1}})^2$~K.
The free-free emissivity of the shocked plasma emitted at frequency
$\nu \ll k_{\rm B}T_{\rm d}/h \sim 480 (v_{\rm sh}/1000\,{\rm km\,s^{-1}})^2$~GHz is
\begin{eqnarray}\label{emiss_ff}
\frac{\epsilon_{\rm ff,\nu}}{\rm erg\,cm^{-3}s^{-1}Hz^{-1}} &\approx& 1.4 \times10^{-33} g(\nu,T_{\rm d}) \nonumber \\
&& \times \left(\frac{n_e}{10^4~\rm cm^{-3}}\right)^2
\left(\frac{v_{\rm sh}}{1000~\rm km\,s^{-1}}\right)^{-1}
\end{eqnarray}
\citep{Lang},
where $k_{\rm B}$ and $h$ are the Boltzmann and Planck constants, respectively. The Gaunt factor is $g(\nu,T_{\rm d}) \sim 0.54 \log \Lambda$ and 
\begin{equation}\label{lambda}
\Lambda \approx 1.1\times10^{9} \left(\frac{v_{\rm sh}}{1000\,\rm km\,s^{-1}}\right)^{2}
\left(\frac{\nu}{\rm GHz}\right)^{-1}
\end{equation}
when $2 \pi \nu \gg \omega_{\rm p}$  and $T_{\rm d} > 3.6\times10^5$~K. Here, $\omega_p \sim 5.6\times10^6 \sqrt{n_e/10^4 \rm cm^{-3}}$~rad~s$^{-1}$ is the electron plasma frequency. We note that  $g(\nu,T) \sim 4.3$ when $v_{\rm sh} =1000$~km~s$^{-1}$ and $\nu \sim 10$~GHz. 

On the other hand, the synchrotron emissivity of a source located at distance $d$ is $\epsilon_{\rm synchr,\nu} = 4\pi d^2 S_{\nu}/V_e$, where $S_{\nu}$ is the synchrotron flux measured at frequency $\nu$ and $V_e\sim R_{\rm j}^3$ is the volume of the synchrotron emitter (see Eq.~\ref{emiss_nu}). By imposing $\epsilon_{\rm synchr, \nu} > \epsilon_{\rm ff,\nu}$, we find that the density of thermal electrons downstream of the shock has to be smaller than $n_{\rm ff}$, where 
\begin{equation}\label{n_cond}
\frac{n_{\rm ff}}{\rm cm^{-3}} \approx  1.4\times10^5
\left(\frac{d}{\rm kpc}\right)
\left(\frac{S_{\nu}}{\rm mJy}\right)^{\frac{1}{2}}
\left(\frac{R_{\rm j}}{10^{16} \rm{cm}}\right)^{-\frac{3}{2}}\left(\frac{v_{\rm sh}}{1000 \,\rm km\,s^{-1}}\right)^{\frac{1}{2}} 
\end{equation}
for the non-thermal emission to dominate at frequency $\nu$ \cite[see][]{Henriksen}. In Fig.~\ref{vshock-min} we plot $n_{\rm ff}$ and in Table~\ref{Tab_dens} we list the values of $n_{\rm ff}$ for the case $v_{\rm sh}=1000$~km~s$^{-1}$ and $R_{\rm j}=R/2$. We can see  that molecular clouds with densities $\sim 10^{5}$~cm$^{-3}$ will not (or marginally) satisfy the condition $n_{\rm mc}<n_{\rm ff}$ for hosting the synchrotron emitter in the shocked region downstream of the bow shock. On the other hand, $n_{\dot M} \ll n_{\rm ff}$ for all the sources in the sample. 

\begin{table*}
\caption{
Observed and derived parameters of non-thermal lobes associated with MYSOs. From left to right we list the name of the source and the non-thermal component in the jet, the observed frequency $\nu$, the measured flux density $S_{\nu}$, the radio spectral index $\alpha$, the major ($\theta_{\rm maj}$) and minor axis ($\theta_{\rm min}$) deconvolved dimensions of the component and its average linear size $R$, the synchrotron emissivity $\epsilon_{\rm synchr}$, the lower limit on the shock velocity $v_{\rm sh,ad}$, the density $n_{\dot M}$, the upper limit on the lobe density given by Eq.~(\ref{n_cond}), and the average density $\langle n_i\rangle$.
We define $R = (R_{\rm maj} + R_{\rm min})/2$ as the characteristic size of the hotspot, where $R_{\rm min,maj} = 4.8\times10^{-6} d\,\theta_{\rm min,maj}$ and $d$ is the  source distance given in Table \ref{Tab_0}.}
\begin{tabular}{lll|cccccc|cccc}
\hline\hline
\multicolumn{3}{c|}{Source} &$\nu$ &$S_{\nu}$ &$\alpha$ &$\theta_{\rm maj}\times\theta_{\rm min}$ &$R$ &$\epsilon_{\rm synchr}$ & $v_{\rm sh,ad}$ &$n_{\dot M}$ &$n_{\rm ff}$ &$\langle n_{\rm i}\rangle$  \\
&& &[GHz] &[mJy]  && [arcsec$^2$] & [cm] &[erg~cm$^{-3}$s$^{-1}$Hz$^{-1}$]& [km~s$^{-1}$] &[cm$^{-3}$] &[cm$^{-3}$] &[cm$^{-3}$]   \\
 \hline
(1)& G263.7434 & N &  9 & 0.56 &$0.45$ & $1.23\times0.36$ & $4.0\times 10^{15}$ &  $4.7\times10^{-30}$& 317 &  $1.0\times10^3$ & $1.6\times10^5$ & $1.2\times10^4$  \\
(2)& G263.7759 &NW &  17 & 2.16&$0.78$ & $1.64\times0.34$ & $5.0\times10^{15}$&  $9.4\times10^{-30}$&353& $1.3\times10^3$ & $2.2\times10^5$& $1.7\times10^4$   \\
(3)&G310.1420 &A4 & 9 & 0.93&$0.70$ & $1.10\times0.89$ &$3.3\times10^{16}$&  $8.4\times10^{-31}$& 373&$2.5\times10^2$ & $6.6\times10^4$&$4.1\times10^4$ \\
(4)& &D &9 & 1.93&$0.46$ & $1.38\times1.05$ &$4.8\times10^{16}$&  $1.2\times10^{-30}$& 377& $2.7\times10^2$ & $8.0\times10^5$&$4.7\times10^4$  \\
(5)& G313.7654& A2 &  9 & 0.14&$0.68$ & $0.85\times0.38$ &$6.6\times10^{16}$&  $2.3\times10^{-31}$&495& $8.6\times10^2$ & $3.4\times10^4$& $5.4\times10^3$  \\
(6) &  & D & 5.5 & 0.15&$0.32$ & $0.80\times0.41$ &$3.4\times10^{16}$&  $2.6\times10^{-31}$ & 497& $8.9\times10^2$ & $3.5\times10^4$&$5.6\times10^3$\\
(7)& G339.8838& NE  & 9 & 2.36&$0.39$ & $0.47\times0.38$ &$8.1\times10^{15}$& $1.3\times10^{-28}$&842&$7.9\times10^4$&$8.5\times10^5$& $2.6\times10^5$  \\
(8) & & SW &   9 & 1.50&$0.72$ & $5.33\times0.85$ &$6.5\times10^{16}$&  $5.6\times10^{-32}$& 463& $3.7\times10^2$ & $1.7\times10^4$&$2.5\times10^3$   \\
(9) &  G343.1261\tnote{*}& N4 &  17 & 1.80&$0.67$ & $0.65\times0.35$ &$1.0\times10^{16}$&  $3.2\times10^{-29}$&674 & $1.6\times10^4$ & $4.2\times10^5$& $8.2\times10^4$ \\
(10) & & S1 &17 & 4.72&$0.45$ & $0.64\times0.35$ &$1.0\times10^{16}$&  $3.8\times10^{-29}$&633& $9.2\times10^3$ & $4.6\times10^5$& $6.5\times10^4$   \\
(11) & G114.0835& B &  1.5 & 0.21&$0.42$ & $2.50\times1.20$ &$1.1\times10^{17}$&  $2.3\times10^{-32}$&271& $0.3\times10^2$ & $1.0\times10^4$&$6.1\times10^2$   \\
\hline
\multicolumn{3}{c|}{Average values}&&&0.55&$1.46\times0.57$&$3.3\times10^{16}$&$1.8\times10^{-30}$&455&$1.2\times10^3$&$1.0\times10^5$&$1.6\times10^{4}$\\
\hline
&&&\multicolumn{6}{c|}{Observational data}&\multicolumn{4}{c}{Section~\ref{Sec_shocks-dens}}\\
\hline\hline
\label{Tab_dens}
\end{tabular}
\end{table*}

Hereafter, we will consider light jets ($\chi < 1$) for which the synchrotron radiation dominates over the free-free emission and  identify the reverse shock  with jet speed ($v_{\rm j}=v_{\rm rs}$). In order to simplify the notation we will name it $v_{\rm sh}$ and use $1000$~km~s$^{-1}$ as a characteristic value\footnote{This value probably overestimates the effective $v_{\rm rs}$ but, on the other hand, $v_{\rm j}$ deduced from proper motions is underestimated given that it corresponds to the velocity projected on the plane of the sky.}. The jet ion density will be $n_{\dot M}\le n_{\rm i}\le n_{\rm ff}$ with a mean value $\langle n_{\rm i} \rangle= \sqrt{n_{\dot M} n_{\rm ff}}$ (see Table~\ref{Tab_dens}).
We note that $\langle n_{\rm i} \rangle$ is similar to 
$n_{\dot M}$ when $R_{\rm j}=R/6$.
In Fig.~\ref{sketch} we show a sketch of the scenario considered in this study. 

Setting constraints on $n_{\rm i}$ and $v_{\rm sh}$ is highly important in our study given that the jet kinetic energy density 
\begin{equation}
\label{Ukin}
\frac{U_{\rm kin}}{\rm erg\,cm^{-3}} 
\simeq \frac{1.7\times10^{-4}}{X_{\rm u}} \left(\frac{n_{\rm i}}{10^4 \rm cm^{-3}}\right)
\left(\frac{v_{\rm sh}}{1000\,\rm km\,s^{-1}}\right)^{2}, 
\end{equation}
and the jet kinetic luminosity
\begin{equation}
\label{Lkin}
\frac{L_{\rm kin}}{\rm erg\,s^{-1}} 
\simeq \frac{2.1\times10^{37}}{X_{\rm u}} \left(\frac{n_{\rm i}}{10^4 \rm cm^{-3}}\right)
\left(\frac{v_{\rm sh}}{1000\,\rm km\,s^{-1}}\right)^{3}
\left(\frac{R_{\rm j}}{10^{16}\rm cm}\right)^2, 
\end{equation}
represent the energy budget to accelerate particles at the shock.
Another very important parameter in our study is the magnetic field.  

\subsection{Magnetic fields}
\label{S:MF}
Maser emission provides information about the orientation and strength of magnetic fields in the molecular outflows associated to the central jet. \citet{Goddi:17} derived magnetic field strengths between 100 and 300 mG in the protostellar jet W3(H$_2$O), which likely results from strong gas compression behind shocks associated with the outflow expansion in the ambient molecular cloud. However, it is difficult to infer the magnetic field orientation and strength in the central jet itself where the origin of the magnetic field is more likely connected to the ejection process from the accretion disk. \cite{Lee_18} reported on the detection of SiO line polarization in the HH~211 protostellar jet, with an estimated magnetic field of about 15~mG at $\sim$300~AU from the central protostar.

Theoretical studies of magnetically-accelerated jets require the action of a poloidal component to accelerate the plasma and a toroidal component to confine it \citep{Casse02}. The toroidal component evolves with the distance $z$ from the star along the jet as $B_{\rm \phi} \propto 1/z$. At the jet base, on the edge of the disk, typical values of the magnetic field strength are obtained from the local beta plasma parameter $\beta_{\rm p} = P_{\rm g}/P_{\rm m}$, where $P_{\rm g}$ and $P_{\rm m}$ are the gas and magnetic pressures, respectively. Jet launching requires $\beta_{\rm p} \ge 1$ \citep{Casse02} giving an upper limit for the magnetic field strength at the base of the jet of $0.3(n_{\rm disc}/10^{10}~{\rm cm}^{-3})(T_{\rm disc}/10^3~{\rm K})~$G where $n_{\rm disc}$ and $T_{\rm disc}$ are the gas density and temperature in the disc, respectively.

Using the model of \citet{Combet08} we can infer typical values of the vertical magnetic field strength in jet emitting disks $B_{\rm z} \sim 0.1$--$10$~G at 1 AU from the star depending on the central mass of the object and the accretion rate. 
We will henceforth consider typical magnetic field strengths at the base of the jet in the range 10~mG--10~G to be conservative. 
By using a $1/z$ dilution factor, we obtain typical magnetic field strengths of $B_{\rm j} \sim 10~\mu\rm{G}$--$10$~mG at distances of the jet termination shock of $\sim 10^3$~AU or ten times smaller at a distance of $10^4$ AU. By considering the flux-freezing condition, \cite{Hartigan_07} showed that the magnetic field strength of variable MHD jets can be described by the relation $(B_{\rm j}/15~\mu{\rm G})\sim (n_i/100\,\rm cm^{-3})^{0.85}$ giving values in agreement with the above estimate. However, at distances larger than $\sim 10^3$~AU from the protostar, densities and magnetic fields are mostly influenced by shocks and rarefaction waves. 

The magnetic field topology at the jet termination shock in high-mass protostars is difficult to assess, but it is likely helical due to a combination of a toroidal and a poloidal component \citep{Cerqueira:97, Cecere:16}, hence the orientation of the magnetic field with respect to the normal of the termination shock front is likely oblique. On the other hand, numerical studies of non-relativistic and magnetically driven jets show that kink instabilities destroy the ordered helical structure of the magnetic field. \citet{Moll_09} showed that at $z \gtrsim 2$--$15 R_{\rm A}$, the toroidal field is dissipated and the poloidal component dominates. Here, $R_{\rm A}$ is the Alfv\'en radius, where the poloidal jet and the Alfv\'en speeds are equal. For the objects of interest, this length scale is typically less than a few tens of AU \citep{Pelletier92}. A dominant poloidal field is also supported by laboratory experiments \citep{Albertazzi_14}.

In this work, acknowledging for the above uncertainties, we adopt  $B_{\rm j} = 10~\mu$G
as a characteristic value of the jet magnetic field in the termination region.
The Alfv\'en speed in the jet is given by 
\begin{equation}\label{v_Alfven}
\frac{v_{\rm A}}{\rm km\,s^{-1}} \simeq 0.2 
\left(\frac{B_{\rm j}}{10~\mu\rm G}\right)
\left(\frac{n_{\rm i}}{10^4~\rm cm^{-3}}\right)^{-\frac{1}{2}},
\end{equation}
and the Alfv\'en Mach number is
\begin{equation}\label{M_Alfven}
M_{\rm A} = \frac{v_{\rm sh}}{v_{\rm A}} \simeq 4800
\left(\frac{v_{\rm sh}}{1000~\rm km\,s^{-1}}\right) 
\left(\frac{B_{\rm j}}{10~\mu\rm G}\right)^{-1}
\left(\frac{n_{\rm i}}{10^4~\rm cm^{-3}}\right)^{\frac{1}{2}}.
\end{equation}
To keep the calculations simple, we derive most of estimates in the parallel shock configuration (i.e. with a purely poloidal magnetic field), but we discuss in Sect.~\ref{S:Obl} 
the impact of the magnetic field obliquity on our results.

\section{Relativistic particles and synchrotron radiation}
\label{S:Rel}

The energy distribution of particles downstream of a shock is a Maxwellian (thermal) distribution with a power-law (non-thermal) tail at a high energies. The transition energy between both distributions is unknown. By using hybrid simulations of parallel non-relativistic shocks, \cite{Caprioli_14a} have shown that immediately behind the shock there is a \emph{bridge} of supra-thermal ions smoothly connecting the thermal peak with the  power-law, while far downstream there is  a sharp boundary between thermal and non-thermal protons at $E_{\rm inj, p} \sim 4 E_{\rm th}$, where $E_{\rm th} = m_pv_{\rm sh}^2/2 \sim 5.2(v_{\rm sh}/1000\,{\rm km\,s^{-1}})^2$~keV, in agreement with \cite{Bell_II}. \cite{Park_15} have shown that the typical energy needed for electron injection into the shock acceleration is comparable with $E_{\rm inj, p}$.
Achieving such injection energy $E_{\rm inj}$ is much more difficult for electrons, as their initial momentum is a factor $m_p/m_e$ smaller. In dense and non-completely ionized medium, ionization and Coulomb losses can suppress the acceleration of low-energy particles. Ionization and Coulomb losses are almost constant at energies below 
$E_0 = 10^{-4} m_{\rm p} c^2$ and $E_{\rm th} = 4\times10^{-6}(T/10^4 \rm{K})m_{\rm p}c^2$, respectively, and therefore they cannot quench the acceleration at the injection energy if
\begin{equation}\label{ion-coulomb}
\frac{n_{\rm i}}{\rm cm^{-3}} <
\frac{2.5\times 10^7}{\kappa}
\left(\frac{B_{\rm j}}{10\rm \mu G}\right)
\left(\frac{v_{\rm sh}}{1000\,\rm km\,s^{-1}}\right)^2 \frac{1}{\mathcal{F}},
\end{equation}
where $\kappa \ge 1$ indicates the departure from the Bohm diffusion regime,  $\mathcal{F} = \max [X_{\rm u}/\sqrt{T_{\rm u,4}},(1-X_{\rm u})]$, and  $T_{\rm u,4}=T_{\rm u}/10^4$~K \citep{Drury_damping}.  For typical parameters in  jets from high mass protostars, and $X_{\rm u} > 0.1$ neither ionization nor Coulomb losses quench the acceleration at any energy.  We consider that the spectrum of accelerated particles is given by Eq.~(\ref{spectrum}).

\subsection{Non-thermal electrons}\label{S:NTE}

Synchrotron emission at frequency $\nu$ in a magnetic field $B_{\rm s}$ is mostly produced by electrons with energy 
\begin{equation}\label{Eq:EE}
    {E_{\rm e} \over \rm{MeV}} \simeq 475 \left({\nu \over \rm{GHz}}\right)^{1/2} \left({B_{\rm s} \over \rm{mG}}\right)^{-1/2},
\end{equation}
indicating that synchrotron photons in the range 100~MHz-10~GHz correspond to relativistic electrons with energies  in the range 150 MeV-1.5 GeV, when $B_{\rm s}=1$~mG. 
The energy density in non-thermal electrons following a power-law distribution $N_{\rm e} =  K_{\rm e} E_{\rm e}^{-s}$ is $U_{\rm e, tot} = K_{\rm e} f_{\rm e}$, where $f_{\rm e}$ is defined in Eq.~(\ref{f_gamma}) and $s = 2\alpha + 1$. For the present study we select non-thermal lobes with $0.3 < \alpha < 0.8$ and then $1.6 < s < 2.6$. The normalization factor $K_{\rm e}$ can be determined from the measured synchrotron flux $S_{\nu}$ at a particular frequency $\nu$, as described in Appendix~\ref{app-1}. By combining Eqs.~(\ref{Ke_ap}) and (\ref{U_gamma}) we find that 
\begin{eqnarray}\label{Ue_eq}
\frac{U_{e,\rm tot}}{\rm erg \,cm^{-3}} &\approx& 4.8 \times10^{-8} \xi_{K}(s)\left(\frac{f_e}{10}\right) \left(\frac{\nu}{\rm GHz}\right)^{\frac{s-1}{2}}\times 
\\ \nonumber
&& 
\left(\frac{\epsilon_{\rm syn, \nu}}{10^{-30}~\rm erg~s^{-1}~cm^{-3} Hz^{-1}}\right)
\left(\frac{B_{\rm s}}{\rm mG}\right)^{-\frac{(s+1)}{2}}, 
\end{eqnarray}
where $\xi_{K}(s)$ is given in Eq.~(\ref{g_s}) and $B_{\rm s}$ is the magnetic field in the synchrotron emitter, i.e. mostly coming from the shock downstream medium\footnote{Depending of the properties of the turbulence around the shock, the magnetic field strength in the synchrotron emitting region has to account for the residence times of the particles downstream and upstream. The exact expression is given by Eq.~(18) in \citet{Parizot06}.}. 
The magnitude of $B_{\rm s}$ is unknown.  It is commonly assumed that $B_{s}$ is in equipartition with non-thermal particles, satisfying  the minimum energy requirement to explain the synchrotron emission. By setting $U_{\rm e, tot} = B_{\rm s}^2/(8\pi)$ we find that the magnetic field in equipartition with relativistic electrons is 
\begin{eqnarray}\label{Beq}
\frac{B_{\rm eq, e}}{\rm mG} &\approx& \xi_{\rm eq}(s)
\left(\frac{f_e}{10}\right)^{\frac{2}{s+5}} \times  \nonumber \\
&& \left(\frac{\epsilon_{\rm syn,\nu}}{10^{-30}~\rm erg~s^{-1}~cm^{-3}~Hz^{-1}}\right)^{\frac{2}{s+5}}
\left(\frac{\nu}{\rm GHz}\right)^{\frac{s-1}{s+5}},
\end{eqnarray}
where $\xi_{\rm eq}(s) \simeq \left(1.2 \xi_{\rm K}(s)\right)^{2/(s+5)}$ is plotted in Fig.~\ref{Fig_zeta}.

\begin{figure}
\includegraphics[width=\columnwidth]{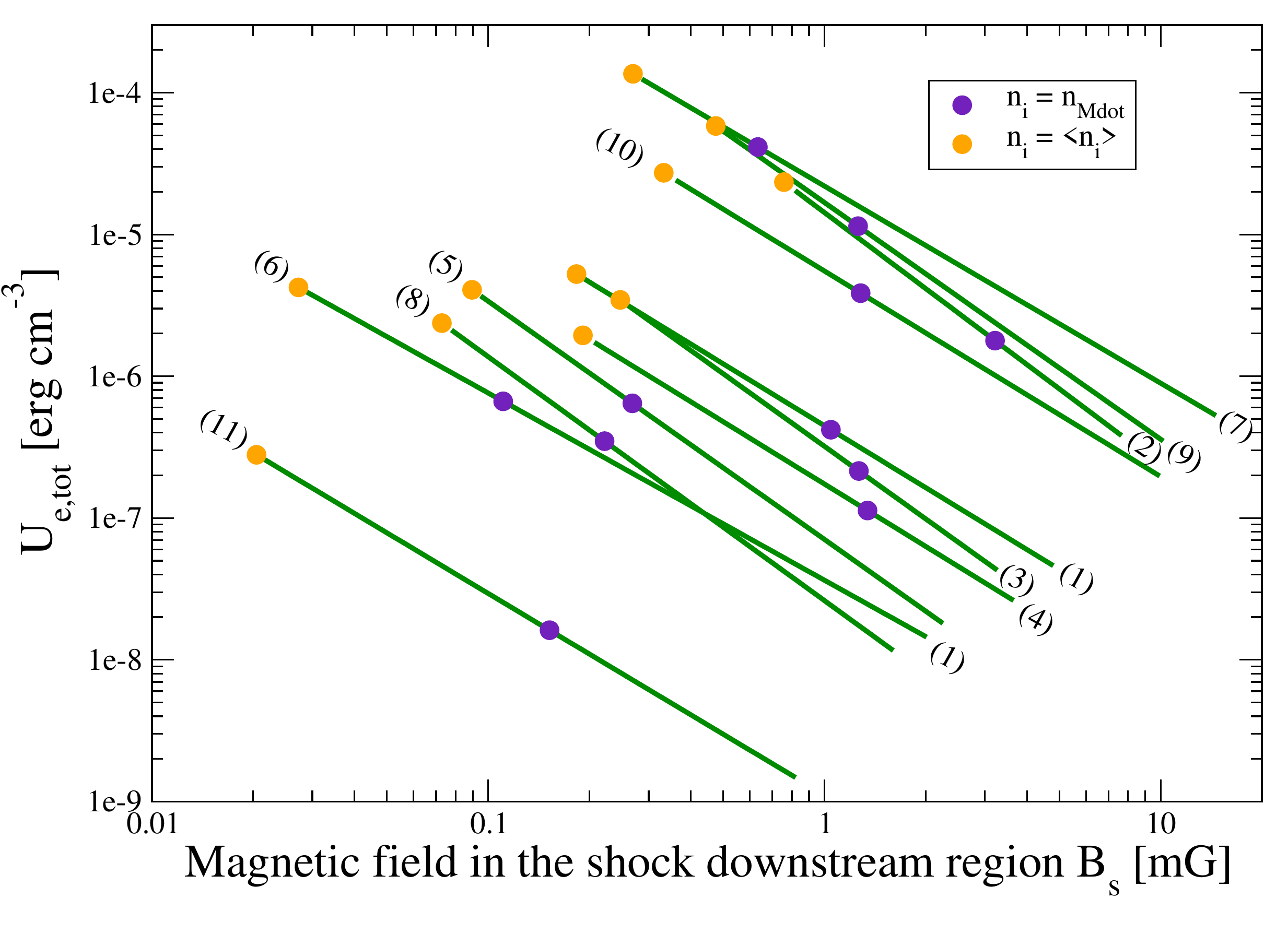}
\caption{Non-thermal electron energy density ($U_{e,\rm tot}$) needed to explain the synchrotron flux at the observed frequency $\nu$ in the sources listed in Table~\ref{T:tab1}. 
The magnetic field strength in the synchrotron emitter is $B_{\rm s} \le B_{\rm eq}$. Orange and violet circles indicate the values of $U_{e,\rm tot}$ and $B_{\rm s}$ that satisfy the condition $U_{\rm nt}=U_{\rm kin}$.
\label{B-Ue}}
\end{figure}

\begin{table*}
\caption{From left to right we list {the source names}, the relativistic electrons and protons energy power-law slope $s = 2 \alpha + 1$, the total protons-to-electrons energy density ratio ($a$, for the case $E_{e,\rm max}=E_{p,\rm max}=1$~TeV), the equipartition magnetic field ($B_{\rm eq}$), the amplified field in the shock downstream region ($B_{\rm sat,NR}$), the total energy density in protons ($U_{p,{\rm tot}}$) and their acceleration efficiency ($\eta_{p,{\rm tot}}$), 
the protons  maximum energy ($E_{p,{\rm max}}$), and the electrons ($K_e$) and protons ($K_p$) normalization constants.  We note that $\eta_{p,{\rm tot}}$ and $E_{p,{\rm max}}$ are computed for 
$n_i$=$n_{\dot M}$ (Eq. \ref{dens_z}) and $n_i$=$\langle n_i\rangle = \sqrt{n_{\dot M}n_{\rm ff}}$ (Eqs.\ref{dens_z},\ref{n_cond}).}
\begin{tabular}{lll|ccc|cccc|cccc}
\hline\hline
\multicolumn{3}{c|}{Source}& $s$ & $a$ &$B_{\rm eq}$ &$B_{\rm sat,NR}$& $U_{p,\rm tot}$ & \multicolumn{2}{c|}{$\eta_{p,\rm tot}$} & 
\multicolumn{2}{c}{$E_{p,\rm max}$}  & $K_e$ & $K_p$\\ 
 & & &  & & [mG]& [mG] & [erg~cm$^{-3}$] &  &   & \multicolumn{2}{c}{[TeV]}& [erg$^{s-1}$~cm$^{-3}$] & [erg$^{s-1}$~cm$^{-3}$]\\
& & &  & & &  &  & $n_i$=$n_{\dot M}$ &  $n_i$=$\langle n_i\rangle$  & $n_i$=$n_{\dot M_i}$ &  $n_i$=$\langle n_i\rangle$ &  &  \\
\hline
(1)& G263.74 &N &1.90 & 18.81 &  4.79& 0.55 & $3.6\times10^{-6}$&0.43& 0.07 & 
0.13 & 0.04& $2.5\times10^{-8}$ & $7.5\times10^{-7}$ \\
(2)&G263.7759&NW &2.56 & 5.11 &  7.70& 0.92 & $1.0\times10^{-5}$&0.93& 0.08 & 
0.13&0.06 &$4.3\times10^{-10}$ & $1.5\times10^{-7}$\\
(3)&G310.1420&A4 &2.40 & 8.90 &  3.27& 0.40&$1.8\times10^{-6}$&0.86& 0.06 & 
0.38&0.14&$3.1\times10^{-10}$ & $5.9\times10^{-8}$\\
(4)&&D &1.92 & 19.24 & 3.65& 0.42&$2.1\times10^{-6}$&0.91&0.10 & 
0.88&0.21& $1.3\times10^{-8}$ & $4.1\times10^{-7}$\\
(5)&G313.7654&A2 &2.36 & 10.14 &2.25 & 0.28& $8.7\times10^{-7}$&0.12&0.02 & 
0.16&0.08& $2.0\times10^{-10}$ & $3.3\times10^{-8}$\\
(6)&&D & 1.64 & 10.14 & 2.01& 0.22 &$5.7\times10^{-7}$&0.08&0.06 & 
0.65&0.26& $2.0\times10^{-8}$ & $2.3\times10^{-7}$\\
(7)&G339.8838&NE &1.78 & 15.06 &  14.6& 1.63&$3.2\times10^{-5}$&0.05&0.04 & 
0.28&0.15&$4.9\times10^{-7}$ & $9.2\times10^{-6}$\\
(8)&&SW &2.43 & 7.80  & 1.60& 0.19 & $4.4\times10^{-7}$&0.14&0.02& 
0.17&0.09&$5.2\times10^{-11}$ & $1.2\times10^{-8}$\\
(9)&G343.1261&N4 &2.34 & 10.80&  10.2&1.22 &$1.7\times10^{-5}$&0.13&0.03 & 
0.17&0.09&$4.8\times10^{-9}$ & $7.4\times10^{-7}$\\
(10)&&S1 &1.89 & 63.07 & 9.93& 1.13 &$1.5\times10^{-5}$&0.20&0.06& 
0.43&0.16&$1.1\times10^{-7}$ & $3.2\times10^{-6}$\\
(11)&G114.0853&B &1.84 & 17.13 &  0.82 & 0.09 &$1.0\times10^{-7}$&0.35&0.05& 
0.43&0.10&$1.1\times10^{-9}$ & $2.5\times10^{-8}$\\
\hline
\multicolumn{3}{l|}{Average values}&2.13&&5.52&0.19&$7.7\times10^{-6}$&0.38&0.05&0.55&0.20&$3.8\times10^{-9}$ & $2.4\times10^{-7}$\\
\hline
&&&\multicolumn{3}{c|}{Section~\ref{S:Rel}}&\multicolumn{4}{c|}{Section~\ref{sec-MFA}}&\multicolumn{2}{c}{Section~\ref{sec-maxE}}&\multicolumn{2}{c}{Section~\ref{Sec_gamma}}\\
\hline\hline
\label{T:tab1}
\end{tabular}
\end{table*}

\subsection{Non-thermal protons}
\label{Sec_equip_protons}

The energy density in  protons with energy $E_{\rm p} \ge m_{\rm i} c^2$ is $U_{p,\rm tot} = a U_{e, \rm tot}$, where $a = (m_{\rm p}/m_{\rm e})^{(s-1)/2}f_{\rm p}/f_{\rm e}$ (see Appendix~\ref{app-2}).
The non-thermal energy density is $U_{\rm nt} = U_{\rm p, tot} + U_{\rm e, tot} =  (1 + a) U_{\rm e, tot}$, and then the magnetic field in equipartition with non-thermal electrons and protons is 
\begin{equation}\label{Beq_a}
B_{\rm eq} = (1+a)^{\frac{1}{2}} B_{\rm eq,e}.
\end{equation}
In Table~\ref{T:tab1} we list the  values of $s$, $a$, and $B_{\rm eq}$ for all the  sources in Table~\ref{Tab_dens} and considering $E_{e,\rm max}=0.1$~TeV in $f_e$. However, we note that $B_{\rm eq}\propto f_e^{2/(s+5)}$ and therefore the dependence on $E_{e,\rm max}$ is negligible. We also point out that $B_{\rm eq}$ is an upper limit for the magnetic field in the synchrotron emitter. In Fig.~\ref{B-Ue} we plot $U_{e,\rm tot}$ for  $B_{\rm s}\le B_{\rm eq}$. Orange and violet circles corresponds to the case where the total acceleration efficiency in non-thermal electrons and protons is $U_{\rm nt}/U_{\rm kin}=1$ and we fixed $v_{\rm sh}=1000 $~km~s$^{-1}$ and $n_{i}=\langle n_{\rm i}\rangle$ and $n_{\dot M}$, respectively, to compute $U_{\rm kin}$.

We define the total proton acceleration efficiency $\eta_{\rm p, tot}=U_{\rm p, tot}/U_{\rm kin}$. These protons can excite various types of instabilities at different scales \cite[see e.g.][for a review]{Alexandre_review}. In particular they drive a current that combined with small perturbations present in the magnetic field can excite NR waves \citep{Tony_04}. Previous studies of DSA in protostellar jets consider that particles diffuse back and forth the shock due to (resonant) Alfv\'en waves \cite[e.g.][]{Crusius-Watzel, Henriksen, IRAS_07, Valenti_10, Marco_15}. However,
we will see that, if the jet (unperturbed) magnetic field $B_{\rm j}$ is smaller than a certain value and the driving parameter (related to the CR current in the upstream medium) $\zeta = 2 \eta_p v_{\rm sh}/c$ is high enough, the dominant instability is non-resonant. The acceleration efficiency $\eta_{\rm p}$ of protons with particular energy $E_{\rm p}$ and energy density $U_p$ in Eq.~(\ref{U_gamma}) is defined as 
$\eta_p=U_p/U_{\rm kin}=\eta_{p,\rm tot}/f_p$, where $f_p$ is defined in Eq.~(\ref{f_gamma}) and plotted in Fig.~\ref{Fig_fs}.

\section{Cosmic-ray streaming instabilities}
\label{streaming}
\begin{figure}
\includegraphics[width=\columnwidth]{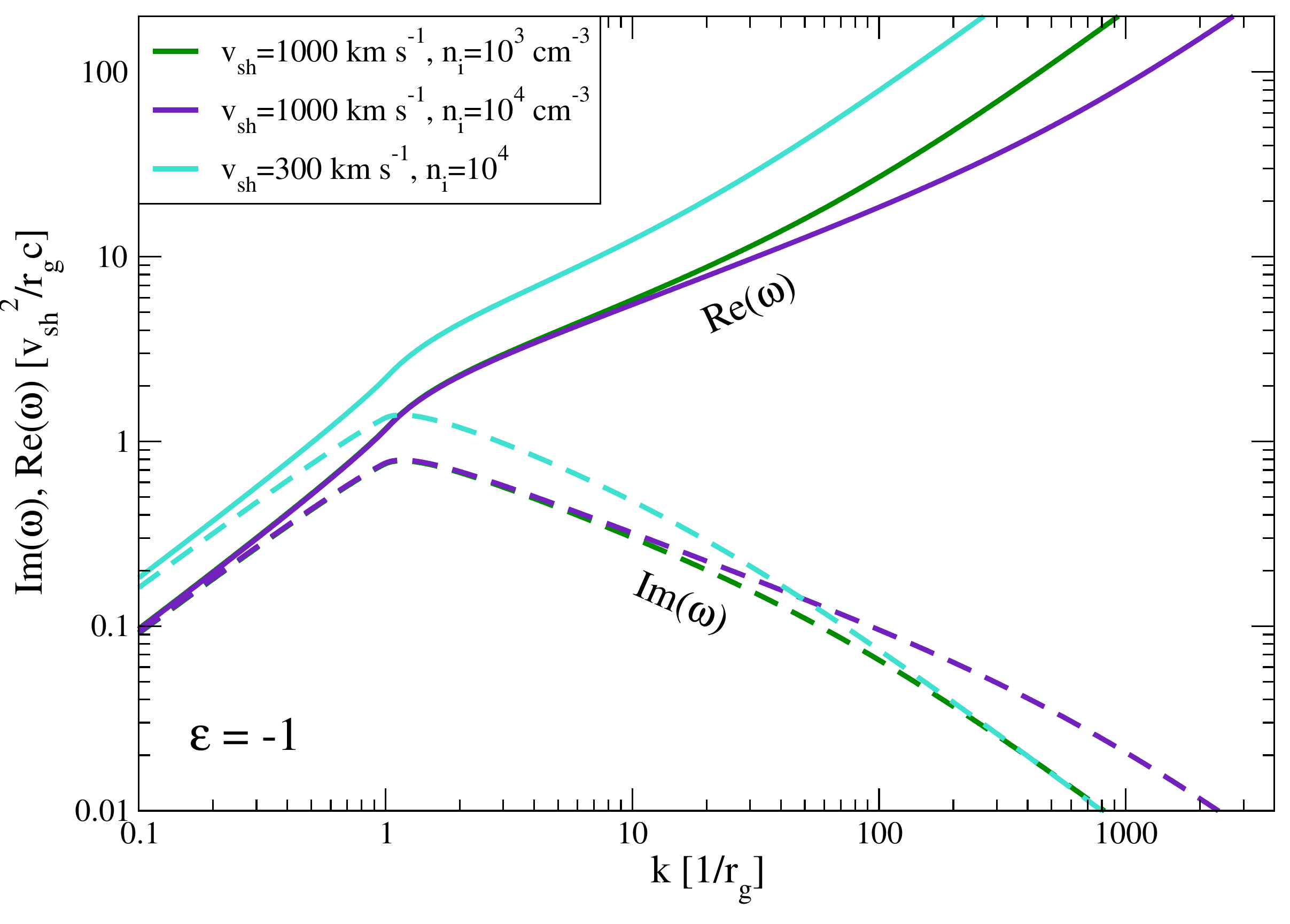}
\includegraphics[width=\columnwidth]{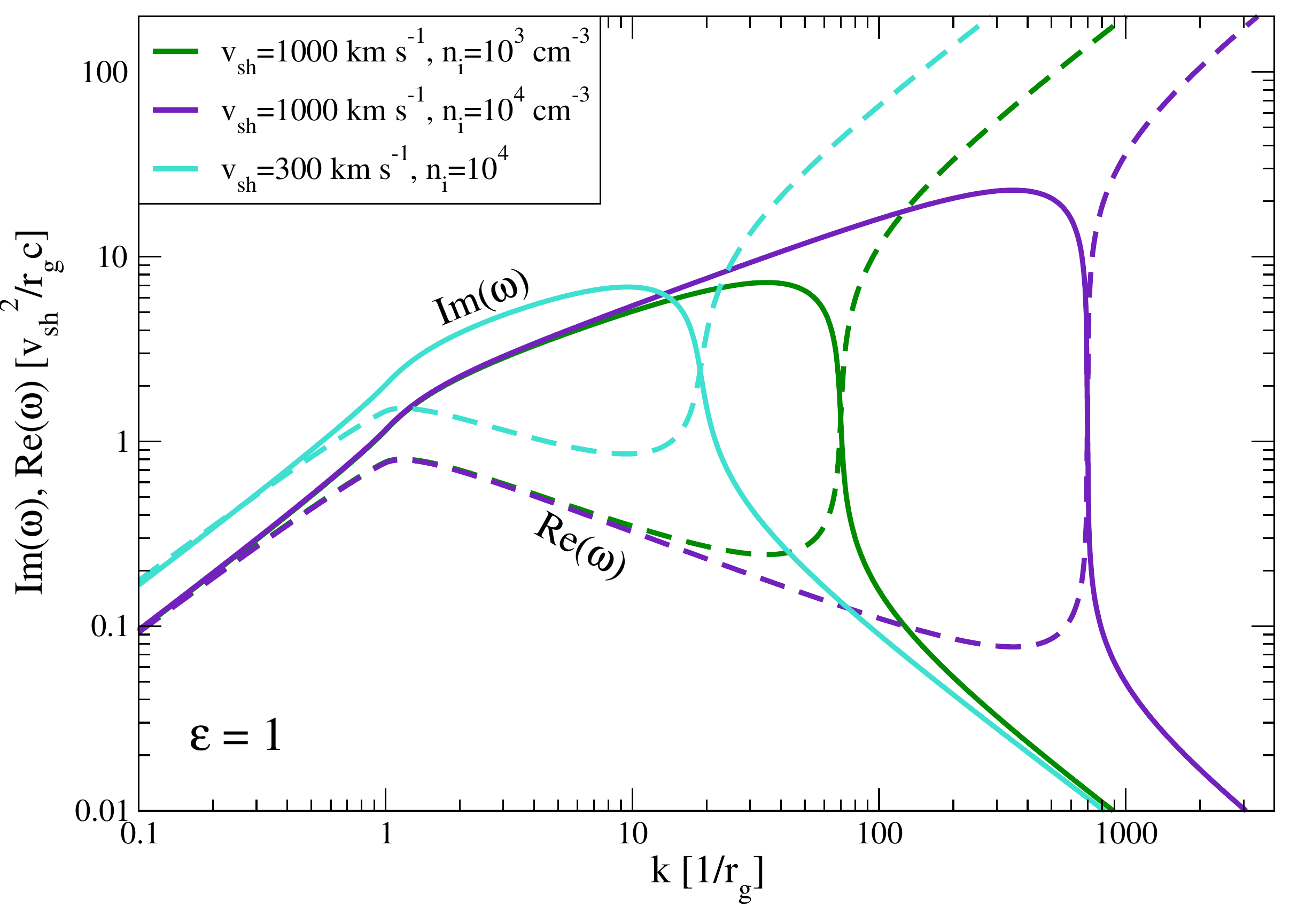}
\caption{Dispersion relation $\omega(k)$ for the resonant (upper and lower panels for $kr_{\rm gm} \le 1$) and non-resonant (lower panel only for $kr_{\rm gm} \ge 1$) modes destabilized by CRs following a power-law energy distribution with $s = 2$. Re$(\omega)$ -- dashed lines -- and Im$(\omega)$ -- solid lines -- are plotted for $v_{\rm sh} = 300$ and $1000$~km~s$^{-1}$ and $n_i = 10^3$ and $10^4$~cm$^{-3}$.
We fix  $B_{\rm j} = 10$~$\mu$G and $\eta_p = 0.01$. 
\label{omega-plot}}
\end{figure}

The current $j_{\rm p} = n_{\rm p} e \,v_{\rm sh}$ of relativistic protons with energy $E_{\rm p}$ and a number density $n_{\rm p} = \eta_{\rm p} U_{\rm kin}/E_{p}$ in Eq.~(\ref{n_gamma})
can drive MHD turbulence due to the force ${\bf j_{p}} \times {\bf B}$ added in the momentum equation \citep{Tony_04,Tony_05}. The parameter $\eta_{\rm p}$ represents the fraction of the kinetic energy imparted into protons with energy $E_{\rm p}$ driving the instabilities.
We consider a parallel shock in the $z$-direction with a small perturbation in the plasma. Wave solutions $\propto \exp(ikz - \omega t)$ of the first order MHD perturbed equations lead to the dispersion relation
\begin{equation}\label{dr}
\omega^2 = k^2v_{\rm A}^2 + \epsilon k \zeta 
\frac{v_{\rm sh}^2}{r_{\rm g}}(\sigma(x,s) - 1),
\end{equation}
where $\epsilon = \pm 1$,  
\begin{equation}\label{rgm}
\frac{r_{\rm g}}{\rm cm} \simeq 3.3\times10^{11}
\left(\frac{E_{p}}{{\rm GeV}}\right)
\left(\frac{B_{\rm j}}{10 \mu \rm G}\right)^{-1}
\end{equation}
is the proton Larmor radius  
and $\sigma(x,s) = 0$ when
$x \equiv kr_{\rm g} >>1$
\citep{Bell_rev_14}. 
The solutions of the dispersion relation in Eq.~(\ref{dr}) are plotted in Fig.~\ref{omega-plot} for  $s=2$, $v_{\rm sh} = 300$ and $1000$~km~s$^{-1}$ and $n_{i}=10^3$ and $10^4$~cm$^{-3}$. We plot Im($\omega$) and Re($\omega$) for $\epsilon =1$  (Bell or NR branch, bottom panel) and $\epsilon =-1$ (Alfv\'en or resonant branch, top panel). Growing modes correspond to Im$(\omega)>0$. Im$(\omega)$ and Re$(\omega)$ are normalized to $1/(3t_{\parallel})$, where $t_{\parallel} = r_{\rm g}c/3v_{\rm sh}^2$ is the time it takes a parallel shock to travel through the layer within which protons of speed $\sim c$ are confined and the upstream diffusion coefficient is $\kappa_{\rm Bohm}= r_{\rm g} c/3$ (Bohm diffusion). More generally, this layer is controlled by the upstream diffusion coefficient $\kappa_{\rm u}$ which depends on the shock obliquity and we define the advection time as $t_{\rm adv}= \kappa_{\rm u}/v_{\rm sh}^2$ (see Sect.~\ref{S:Obl}). 

We can see in Fig.~\ref{omega-plot} (top panel) that resonant waves are unstable when $k r_{\rm g} \approx 1$ and therefore this instability is resonant. On the other hand, the bottom panel shows that in the NR regime ($k r_{\rm g} > 1$) $\rm Im(\omega) \gg Re(\omega)$ and then the NR mode is almost purely growing. The maximum growth rate $\Gamma_{\rm max}\equiv \max(\rm Im(\omega))$ moves to the resonant regime ($k r_{\rm g} \approx 1$) when $v_{\rm sh}$ decreases. However, even when $v_{\rm sh}$ is as small as 300~km~s$^{-1}$, the NR mode is still dominant due to the large ion density of protostellar jets. Both modes are present in the plasma, although the dominance of one over the other depends on the competition between the Alfv\'en (that contains $v_{\rm A}$) and the driving (that contains $\zeta$) terms in Eq.~(\ref{dr}). 
We note that the NR mode exists when $1/r_{\rm g} \le k \le k_{\rm max}$, where $k_{\rm max}=\Gamma_{\rm max}/v_{\rm A}$. Therefore, the condition for the growth of the NR mode is $k_{\rm max} r_{\rm g} > 1$, which can be  written as $\zeta M_{\rm A}^2 >1$, where
\begin{equation}\label{Eq:driving_tot}
\zeta M_{\rm A}^2 \simeq 3072
\left(\frac{\eta_{p}}{0.02}\right) 
\left(\frac{n_i}{10^4 \rm cm^{-3}}\right)
\left(\frac{B_{\rm j}}{10\mu\rm G}\right)^{-2}
\left(\frac{v_{\rm sh}}{1000\, \rm km\,s^{-1}}\right)^{3}.
\end{equation}
It is clear from Eq.~(\ref{Eq:driving_tot}) that the NR mode is more easily triggered by the fastest shocks. The condition $\zeta M_{\rm A}^2 >1$ can be written also as $U_p (v_{\rm sh}/c) > B_{\rm j}^2/(8\pi)$, or equivalently 
\begin{equation}\label{Eq:driving_tot_Up}
\frac{U_p}{\rm erg\,cm^{-3}} > 1.2\times 10^{-9}
\left(\frac{B_{\rm j}}{10 \mu\rm G}\right)^{-2}
\left(\frac{v_{\rm sh}}{1000\, \rm km\,s^{-1}}\right)^{-1}.
\end{equation}

In the limit $\zeta M_{\rm A}^2 > 1$, the NR instability dominates. In the regime $k_{\rm max,NR} \ge k \ge 1/r_{\rm g}$, $\sigma \approx 0$ and the dispersion relation in Eq.~(\ref{dr}) reduces to
\begin{equation}\label{dr_Bell}
\omega^2 = k^2v_A^2 - k \zeta \frac{v_{\rm sh}^2}{r_{\rm g}}.
\end{equation}
The maximum growth rate of the NR instability excited by protons with energy $E_{\rm p}$ and carrying a current of strength $j_{\rm p}$ is $\Gamma_{\rm max} = 0.5 (j_{\rm p}/c) \sqrt{\pi/ n_{\rm i}\,m_{\rm i}}$, giving
\begin{eqnarray}\label{gamma_max}
 \frac{\Gamma_{\rm max,NR}}{\rm s^{-1}} &\simeq& 2.6\times10^{-5} \left(\frac{\eta_p}{0.02}\right)
\left(\frac{v_{\rm sh}}{1000\,\rm km\,s^{-1}}\right)^3 \nonumber \\
&&\times \left(\frac{n_{i}}{10^4\,\rm cm^{-3}}\right)^{\frac{1}{2}} \left(\frac{E_{\rm p}}{\rm{GeV}}\right)^{-1} \ .
\end{eqnarray}
We note that $\Gamma_{\rm max,NR}$ is independent of $B_{\rm j}$, however the wavenumber $k_{\rm max,NR} = \Gamma_{\rm max,NR}/v_{\rm A} = \zeta M_{\rm A}^2/r_{\rm g}$ of the fastest growing modes does depend on $B_{\rm j}$ (see Eq.~\ref{Eq:driving_tot}).
We also note that as $\Gamma_{\rm max,NR}$ and $k_{\rm max,NR}$ decrease with $v_{\rm sh}$. Resonant modes should take over as it is the case in low mass YSOs \citep{Marco_16}.

In the limit $\zeta M_A^2 <1$, the Alfv\'en instability dominates and the dispersion relation in Eq.~(\ref{dr}) reduces to $\omega^2 = k^2v_A^2$.
The maximum growth rate of resonant modes occurs for parallel propagating modes \citep{Achterberg81} at $k_{\rm max,R}\approx \sqrt{3}/r_{\rm g}$ and therefore $k_{\rm max,R} r_{\rm g}\approx 1$, as shown in Fig.~\ref{omega-plot} (top panel). 
 The growth rate reaches its maximum value at $\Gamma_{\rm max,R}\sim \Gamma_{\rm max,NR}$
\cite[e.g.][]{Pelletier06,Amato09}.
Alfven instabilities dominates over Bell instabilities
when 
\begin{equation}
\frac{v_{\rm sh}}{\rm km\, s^{-1}} < 128 
\left(\frac{\eta_p}{0.02}\right)^{-\frac{1}{2}}
\left(\frac{B_{\rm j}}{10\mu\rm G}\right)
\left(\frac{n_i}{10^4 \rm cm^{-3}}\right)^{-\frac{1}{2}}.
\end{equation}
This regime was studied by \cite{Marco_15,Marco_16} in the context of low mass protostars, where the jet typical velocities are $\sim 100$~km~s$^{-1}$. 
The important aspect of destabilizing perturbations in resonance with CRs, i.e. when $k r_{\rm g} \approx 1$, is that the scattering process is more efficient. However, this does not guarantee to have a strong amplification. In the present paper we mostly focus on the NR instability.

\subsection{Magnetic field obliquity}\label{S:Obl}
In the perpendicular shock case the wave speed in Eq.~(\ref{dr_Bell}) now includes a contribution of the compression modes characterized by the local sound speed $c_{\rm s}$ \citep{Tony_05, Marcowith:18}. If $c_{\rm s} < v_{\rm A}$ the non-resonant growth rate is unmodified \citep{Tony_05, Matthews:17}. If $c_{\rm s} > v_{\rm A}$ the growth rate drops because of thermal effects (see Section~\ref{S:THR}). 

The necessary condition for the streaming instabilities to grow is also modified. In perpendicular shocks, the precursor size is shorten because the particle transport is controlled by the perpendicular diffusion $\kappa_{\rm u} =\kappa_\perp = \kappa_{\rm Bohm} \kappa/(1+\kappa^2)$ \citep{Forman75} where $\kappa= \kappa_{\parallel}/\kappa_{\rm Bohm}$. If angular diffusion proceeds at a smaller rate than Bohm then $\kappa \gg 1$ and $\kappa_\perp \simeq \kappa_{\rm Bohm}/\kappa$. We  derive the general expression $\Gamma_{\rm max,NR}t_{\rm adv} \sim \eta_{\rm p} (v_{\rm sh}/v_{\rm A}) \kappa^{b}/24$,
where $b=1$ and $-1$ (with $\kappa > 1$) for parallel and perpendicular shocks, respectively. 
In perpendicular shocks, $\Gamma_{\rm max, NR}t_{\rm adv} > 1$ gives $\eta_{p} > \kappa/M_{\rm A}$. High $\kappa$ values probably quench the development of fast streaming instabilities in the configuration of a weakly CR modified shock. In the mean time, if $\kappa > 100$ then $\eta_{p} > 0.1$ and the destabilization of streaming instabilities occurs in the regime of strongly CR modified shocks, a case beyond the scope of the present linear analysis. 

\subsection{Reduction of the NR growth rate due to environmental conditions}

Different effects may reduce the growth rate of the NR instability. 
In this paper we discuss 
thermal effects and  ion-neutral collisions in non-completely ionized jets.

\subsubsection{Thermal effects}
\label{S:THR}

Thermal effects are important when $(v_{\rm A}/v_i)^3<(n_p/n_i)(v_{\rm sh}/v_i)$, where
the ion speed is $v_i^2 = k_{\rm B} T_{\rm u}/m_p$ \citep{Zweibel_10}.
This leads to the condition
\begin{eqnarray}\label{Tu_th}
\frac{T_{\rm u}}{\rm K} &>& 1.1\times10^4
\left(\frac{\eta_{\rm p}}{0.02}\right)^{-1}
\left(\frac{E_p}{\rm GeV}\right)\nonumber \\
&& \times \left(\frac{v_{\rm sh}}{1000\,\rm km\,s^{-1}}\right)^{-3} \left(\frac{B_{\rm j}}{10~\mu \rm G}\right)^3
\left(\frac{n_i}{10^4\rm cm^{-3}}\right)^{-\frac{3}{2}}
\end{eqnarray}
in which case an extra term $\propto k^2 T_{\rm u}^2 r_{\rm g} \omega$ needs to be added in the dispersion relation in Eq.~(\ref{dr}). 
The maximum growth rate of the thermally modified Bell instability is 
\begin{equation}\label{gamma_max_th}
\Gamma_{\rm max, th} \simeq \omega_{\rm ci}
\left(\frac{n_{\rm p}}{n_{\rm i}}\right)^{\frac{2}{3}}
\left(\frac{v_{\rm sh}}{v_i}\right)^{\frac{2}{3}},
\end{equation}
where $\omega_{\rm ci}=eB_{\rm j}/(m_p c)$ is the ion cyclotron frequency \citep{Brian_07}.
For typical values in protostellar jets, 
\begin{eqnarray}\label{gamma_max_th}
\frac{\Gamma_{\rm max, th}}{\Gamma_{\rm max, NR}} &\simeq& 1.9
\left(\frac{\eta_{\rm p}}{0.02}\right)^{-\frac{1}{3}}
\left(\frac{E_p}{\rm GeV}\right)^{\frac{1}{3}}
\left(\frac{T_{\rm u}}{10^4~\rm K}\right)^{-\frac{1}{3}}\nonumber \\
&& \times \left(\frac{v_{\rm sh}}{1000\,\rm km\,s^{-1}}\right)^{-1} \left(\frac{B_{\rm j}}{10~\mu \rm G}\right)
\left(\frac{n_i}{10^4\rm cm^{-3}}\right)^{-\frac{1}{2}}.
\end{eqnarray}
We stress that $\Gamma_{\rm max, th} < \Gamma_{\rm max, NR}$ when the temperature is within the range where the condition in Eq.~\eqref{Tu_th} is satisfied. 
The thermally modified Bell instability is damped when  $\Gamma_{\rm max, th} t_{\rm adv} < 1$, i.e. when
\begin{equation}\label{Tu_min}
\frac{T_{\rm u}}{\rm K} > 
4.5\times10^{7}
\left(\frac{\eta_p}{0.02}\right)^{2}
\left(\frac{E_{p}}{\rm GeV}\right),
\end{equation}
where we have assumed $t_{\rm adv}=t_{\parallel}$. 
When thermal effects are taken into account,
they can reduce the growth rate of the NR instability but we do not expect strong damping by thermal effects over the development of the NR streaming modes, unless the proton acceleration efficiency $\eta_p$ is unreasonably small in which case the NR modes are not destabilized.

\subsubsection{Partially ionized medium}
\label{non-ionized}

\begin{figure}
\includegraphics[width=\columnwidth]{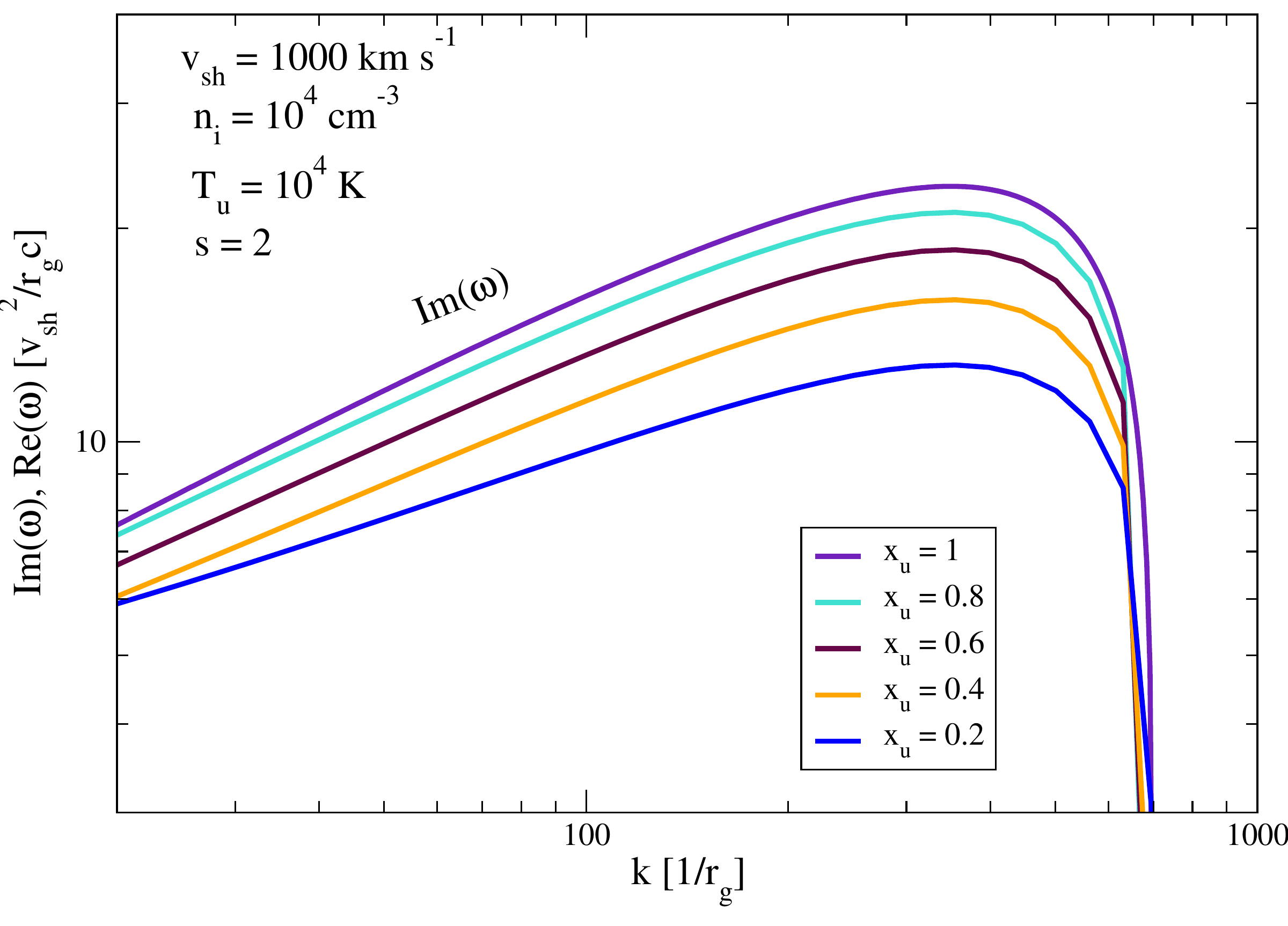}
\caption{Growth rate for different values of the ionization fraction upstream of the shock.  
}
\label{growth}
\end{figure}

In a partially ionized protostellar jet (i.e. $X_{\rm u} < 1$)
the friction arising between charged and neutral particles can  quench the growth of CR driven instabilities at shock precursors, and therefore DSA is less efficient \cite[e.g.][]{Drury_damping, Brian_07}.
When ion-neutral collisions are taken into account in the MHD equations, the dispersion relation of the CR-driven instability becomes 
\begin{eqnarray}
\label{wk_ion}
\omega^3 + i\omega^2 \nu_{\rm in}\left(\frac{1}{1-X_{\rm u}}\right)+ 
\omega \left(\epsilon \zeta \frac{v_{\rm sh}^2}{r_{\rm gm}}  k (1 - \sigma) 
-k^2 v_{\rm A}^2\right)+ && \nonumber \\
i\nu_{\rm in}\left(\frac{X_{\rm u}}{1-X_{\rm u}}\right)\left(\epsilon \zeta \frac{v_{\rm sh}^2}{r_{\rm gm}}  k (1 - \sigma) - k^2v_{\rm A}^2 \right) = 0,
\end{eqnarray}
where the ion-neutral collision frequency is given by 
\begin{equation}
\frac{\nu_{\rm in}}{\rm s^{-1}} \simeq 8.9\times10^{-4} \left(\frac{T_{\rm u}}{10^4~\rm{K}}\right)^{0.5}~\left(\frac{n_{\rm n}}{10^5\rm cm^{-3}}\right),
\end{equation}
in the shock upstream region with $T_{\rm u} > 140$~K and neutral density $n_{\rm n} = (1/X_{\rm u}-1)n_i$ \citep{Jean09}.
We point out that in a completely ionized plasma $X_{\rm u} = 1$, $n_{\rm n} = 0$ and then  $\nu_{\rm in} = 0$. In such a case the dispersion relation in Eq.~(\ref{wk_ion}) is identical to Eq.~(\ref{dr}).
We solve Eq.~(\ref{wk_ion}) for $\epsilon=1$ and different values of $X_{\rm u}$. 
As pointed out by \cite{Brian_07}, ion-neutral collisions are unable to stabilize the Bell modes although the maximum growth rate decreases with $X_{\rm u}$, as we illustrate in Fig.~\ref{growth}. This behaviour for $\Gamma_{\rm max,inc}$ (the maximum growth rate in the incompletely ionized jet) indicates that longer times are required to satisfy the condition $\Gamma_{\rm max,inc} t_{\rm adv} >1$ to excite the NR modes. 

\section{Magnetic field amplification}
\label{sec-MFA}

Magnetic fields in the synchrotron emitter in protostellar jets have strengths $B_{\rm s} \approx 0.1$--$10$~mG (see Sect.~\ref{S:Rel}). These values are larger than the expected magnetic field in the jet termination region, $B_{\rm j} \approx 1$--$100~\mu$G, requiring  amplification of the magnetic field, given that compression by the shock does not produce strong enough fields. In this study we consider the magnetic field amplification by the Bell 
instability discussed in the previous section, but in the non-linear case \citep{Tony_04,Tony_05}.

In the linear regime analyzed in Sect.~\ref{streaming}, the magnetic field increases exponentially with time until it reaches a value $B \sim 2 B_{\rm j}$, after that the amplification enters in a non-linear regime and the magnetic field growth becomes linear with time.
At a particle energy $E_{\rm p}$, the amplified magnetic field saturates at
\begin{equation}\label{B_sat_E}
\frac{B_{\rm p}^2}{8\pi} \sim \eta_p \frac{n_{\rm i} m_{\rm i} v_{\rm sh}^3}{c} = U_p \left(\frac{v_{\rm sh}}{c}\right)
\end{equation}
\citep{Tony_04, Pelletier06,ZirakashviliII_08}.
The above expression refers to magnetic field amplification by protons with energy $E_{\rm p}$. However, the distribution of relativistic protons driving the current spans from $\approx 1$~GeV to $E_{\rm p, max} \approx 1$~TeV (see Sect.~\ref{sec-maxE}) and therefore the (total) saturated magnetic field $B_{\rm sat,NR}$ immediately upstream of the shock is better estimated by considering the total proton population with energy density $U_{p,\rm tot}$ and acceleration efficiency $\eta_{p,\rm tot}$, which gives
\begin{equation}\label{Bsat}
\begin{split}
\frac{B_{\rm sat, NR}}{\rm{mG}} &\approx 
0.3
\left(\frac{U_{p, \rm tot}}{10^{-6}\,\rm erg\,cm^{-3}}\right)^{\frac{1}{2}}
\left(\frac{v_{\rm sh}}{1000\, \rm km\,s^{-1}}\right)^{\frac{1}{2}}\\
& \approx 1.2
\left(\frac{\eta_{p,\rm tot}}{0.1}\right)^{\frac{1}{2}}
\left(\frac{n_i}{10^4\,\rm cm^{-3}}\right)^{\frac{1}{2}} 
\left(\frac{v_{\rm sh}}{1000\, \rm km\,s^{-1}}\right)^{\frac{3}{2}}.
\end{split}
\end{equation}
We note that $B_{\rm sat, NR}$ does not depends on $B_{\rm j}$.

Once the magnetic field strength reaches the saturation value $B_{\rm sat,NR}$, the Alfv\'en velocity becomes larger and $\eta_{\rm p} M_{\rm A}^2$ decreases. At the same time the Larmor radius of particles decreases down to values $r_{\rm g} \sim 1/k$ and then the NR Bell instability becomes subdominant. 
\citet{Caprioli_14b} proposed a refined model for the growth of the non-resonant modes including a non-linear stage (when energy densities in the turbulent and background magnetic fields become similar). They find a maximum magnetic field strength $B_{\rm max}/B_{\rm j} \sim M/\sqrt{2}$, where $M=v_{\rm sh}/c_{\rm s}$ is the sonic Mach number and $c_{\rm s}$ is the sound speed.
In our case we have moderate $M$ with a typical range 5-50, which makes the result obtained in Eq.~(\ref{Bsat}) likely optimistic. It should be also noted that \citet{Caprioli_14b} conduct their study in the regime of beta plasma parameter $\beta_{\rm p} = 2 (c_{\rm s}/v_{\rm A})^2 \sim 1$ whereas MYSO jets have rather $\beta_{\rm p} > 1$ if not $\gg 1$ because they are relatively hot and dense with a weak magnetic field.

\subsection{The magnetic field in the shock downstream region}\label{sec_equip}

The perturbations produced by the NR instability can lead to various effects in the shock downstream region \citep{Giacalone_2007}. As NR modes are not normal modes of the plasma, they are expected to be damped rapidly once the source of excitation disappears.  Weibel instabilities would also contribute to produce magnetic field fluctuations upstream of the shocks. However,  Weibel modes are expected to decay over a few plasma skin depth downstream, so over much smaller scales  than $R_{\rm j}$. Conversely, dynamo processes can further amplify the magnetic field produced at the shock front \citep{Tony_04, Pelletier06, 2009ApJ...695..825I, Marcowith10, vanMarle18, Tzeferacos18}, even if the upstream region is partially ionized \citep{2017ApJ...850..126X}. This aspect involves a model of non-linear evolution of the plasma and it is beyond the scope of this paper. For simplicity, we consider that the amplified magnetic field in the shock upstream region is compressed by the shock and maintained on a spatial scale $\sim R$ (the size of the synchrotron emitter).

\begin{figure}
\includegraphics[width=\columnwidth]{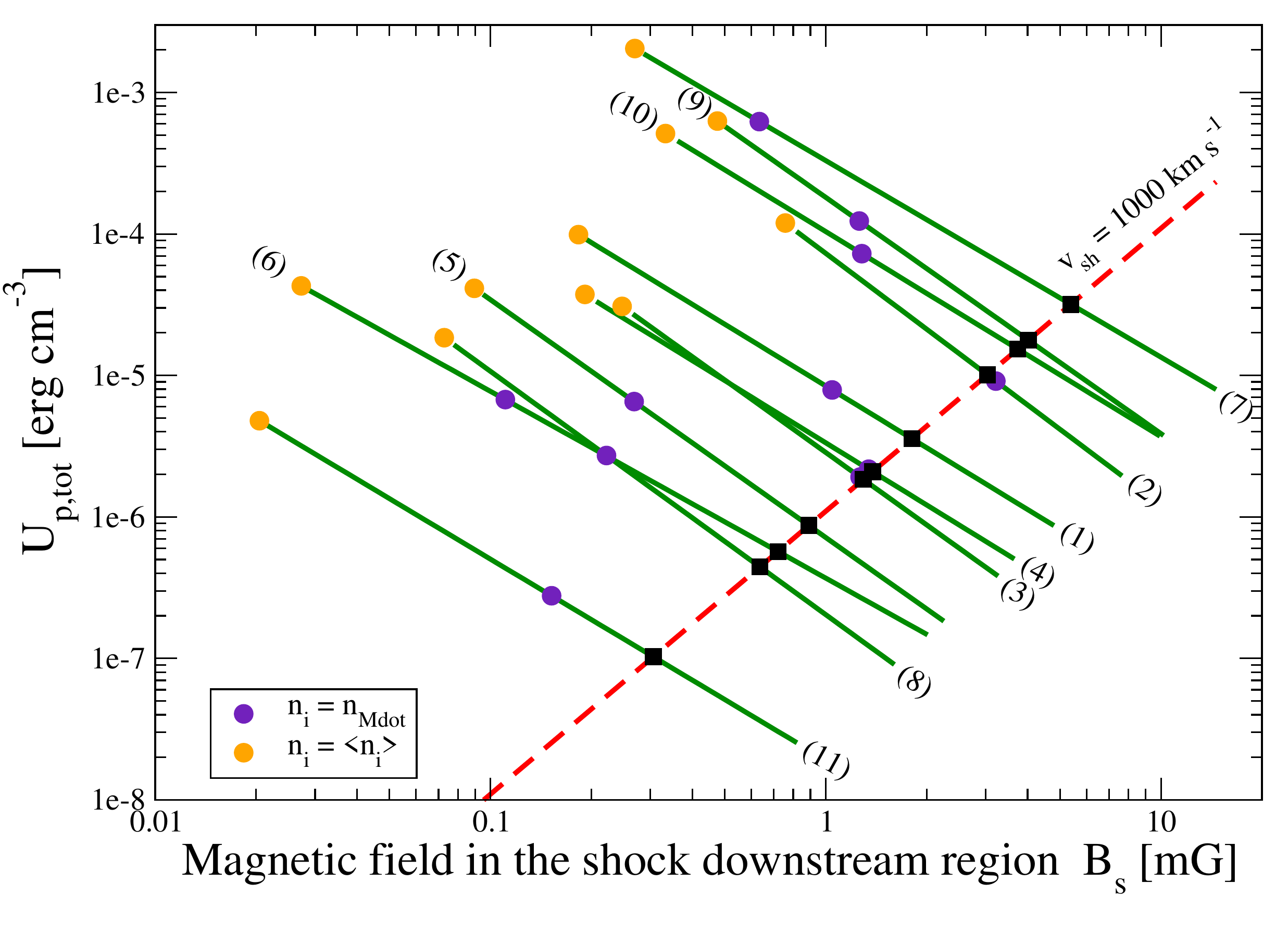}
\caption{Non-thermal protons energy density ($U_{p,\rm tot}=a U_{e,\rm tot}$) needed to explain the synchrotron flux at the observed frequency $\nu$ in the sources listed in Table~\ref{T:tab1} (green-solid lines). 
The magnetic field strength in the synchrotron emitter is
$B_{\rm s} \le B_{\rm eq}$, as in Fig.~\ref{B-Ue}.
The red dashed line represents the energy density in non-thermal protons needed to amplify the magnetic field up to the saturation value $B_{\rm sat,NR}$ as indicated in Eq.~\eqref{Bsat}. Black squares indicate the value of the magnetic field when 
$v_{\rm sh}=1000$~km~s$^{-1}$. Orange and violet circles indicate the values of $U_{p,\rm tot}$ and $B_{\rm s}$ that satisfy the condition $U_{\rm nt}=U_{\rm kin}$.
\label{find_Bs}}
\end{figure}

In the shock downstream region, the (turbulent) isotropic upstream random magnetic field is compressed by the shock by a factor $r_{\rm B}=\sqrt{(2r^2 +1)/3} \approx 3.3$ when the shock compression ratio is $r = 4$ \citep[e.g.][]{Parizot06,ZirakashviliII_08}\footnote{Hereafter we use $r=4$ and $r_{\rm B}=3.3$.}. Therefore, the amplified field downstream of the shock is $B_{\rm sat,d}=r_{\rm B}B_{\rm sat,NR}$. On the other hand, the magnetic field in the synchrotron emitter $B_{\rm s}$, i.e. mostly coming from the shock downstream region, is a function of the energy density in non-thermal electrons.
In Fig.~\ref{find_Bs} we plot $U_{\rm p, tot} = aU_{\rm e, tot}$ (green-solid lines), where $U_{\rm e, tot}$ is given by Eq.~\eqref{Ue_eq}, and $U_{\rm p, tot} = (c/v_{\rm sh})B_{\rm sat,NR}^2/(8\pi)$ by Eq.~\eqref{Bsat} (red-dashed line).
By assuming that $B_{\rm sat,d}=B_{\rm s}$ we find that
\begin{equation}\label{Bsat_eq}
\frac{B_{\rm sat,d}}{B_{\rm eq}} = \left[r_{\rm B}^2\left(\frac{a}{1+a}\right) \left(\frac{v_{\rm
  sh}}{c}\right)\right]^{\frac{2}{s+5}}\sim
0.38 \xi_{\rm sat}(s)\left(\frac{v_{\rm sh}}{1000\,\rm km\,s^{-1}}\right)^{\frac{2}{s+5}}
\end{equation}
and
\begin{equation}
\frac{U_{p,\rm tot}}{\rm erg \,cm^{-3}}=1.7\times10^{-6}\xi_{\rm sat}^2 \left(\frac{v_{\rm sh}}{1000 \rm km\,s^{-1}}\right)^{-\frac{s+1}{s+5}} 
\left(\frac{B_{\rm eq}}{\rm mG}\right)^2,
\end{equation}
where $\xi_{\rm sat} = 10^{0.42(s-2)/(s+5)}$ (see Fig.~\ref{Fig_zeta}) and we have fixed $r_{\rm B}=3.3$ and $a/(1+a)\sim 1$. 
In Fig.~\ref{find_Bs} we indicate $B_{\rm s}$ and $U_{p,\rm tot}$ for the cases with $v_{\rm sh}=1000$~km~s$^{-1}$ (black  squares). In Table~\ref{T:tab1} we list $B_{\rm sat,NR}=B_{\rm s}/3.3$ and $U_{p,\rm tot}$ for $v_{\rm sh}=1000$~km~s$^{-1}$. We list also $\eta_{p,\rm tot}=U_{p,\rm tot}/U_{\rm kin}$ 
for $n_i=n_{\dot M}$ and $\langle n_i\rangle$.
We remark that sources G263.7759~NW (2) and G310.1420~A4,D (3,4) have $\eta_{p,\rm tot}\sim 1$  when $B_{\rm s}=3.3B_{\rm sat,NR}$, $n_i= n_{\dot M}$ and $v_{\rm sh}=1000$~km~s$^{-1}$.
Therefore, we conclude that in these sources the magnetic field either cannot be amplified by Bell instabilities or $n_i > n_{\dot M}$.

Contrary to the case of supernova remnants where the magnetic field in the shock downstream region is calculated from the synchrotron cooling length of X-ray emitting electrons, in the non-thermal lobes considered in the present study the radio emitting electrons are not in the fast cooling regime. We have therefore  estimated the magnetic field in the non-thermal hotspots by assuming that upstream of the shock the magnetic field is amplified by the Bell's instability up the the saturation regime given by Eq.~(\ref{Bsat}) 
and by equating $B_s=3.3 B_{\rm sat,NR}$.
By assuming that the downstream magnetic field is $B_{\rm s}=B_{\rm sat,d}=3.3B_{\rm sat,NR}$ we determine $B_{\rm s}$, $U_{p,\rm tot}$, and $\eta_{p,\rm tot}$. Once we know the
protons acceleration efficiency and the magnetic field,  we compute in the following sections the maximum energy of particles (Sect.~\ref{sec-maxE}) and the $\gamma$-ray emission (Sect.~\ref{Sec_gamma}), respectively.

\section{Maximum energies}
\label{sec-maxE}
\subsection{Protons}
If the non-resonant streaming instability is important and the magnetic field is amplified, the maximum energy of protons is determined by the amount of protons that escape from the shock upstream region \citep{ZirakashviliII_08,Tony_13_Escape}. Given that only the most energetic protons can penetrate far upstream from the shock and amplify the magnetic field in the shock precursor,
the available time to accelerate these particles is $\sim 5/\Gamma_{\rm max,NR}(E_{\rm p, max})$, where $\Gamma_{\rm max,NR}(E_{\rm p, max})$ is the maximum growth rate of NR modes driven by protons with an energy $E_{\rm p, max}$. By equating $5/\Gamma_{\rm max,NR}(E_{p,\rm max}) = R_{\rm j}/v_{\rm sh}$ and using Eq.~(\ref{n_gamma}) with $E_{\rm p}=E_{\rm p, max}$ we find 
\begin{equation}\label{Ep-max}
\frac{E_{p, \rm max}}{\rm m_{\rm p} c^2} = \left\{\begin{array}{ll}
(2-s) F & s<2  \\
 \log\left(\frac{E_{p,\rm max}}{\rm GeV}\right)^{-1} F
&  s=2  \\
\left[(s-2) \frac{1}{m_pc^2} F\right]^{\frac{1}{s-1}} &  s>2
\end{array}\right. 
\end{equation}
\citep{Tony_13_Escape,Klara_Tony_14}, where
\begin{equation}\label{Ep-max2}
F 
\simeq 65.83 \left(\frac{U_{p,\rm tot}}{10^{-5}\rm erg\,cm^{-3}}\right)
\left(\frac{R_{\rm j}}{10^{16}\rm cm}\right)
\left(\frac{n_{i}}{10^4\,\rm cm^{-3}}\right)^{-\frac{1}{2}}.
\end{equation}
 Values of $E_{\rm p, max}$ for the non-thermal lobes in our study are listed in Table~\ref{T:tab1} for the case  $v_{\rm sh}=1000\,{\rm km\,s^{-1}}$ and $R_{\rm j}=R/2$. We also considered $n_{i}=n_{\dot M_i}$ and $\langle n_i \rangle$.  Notice that around the shock the proton mean free path is $\kappa r_{\rm g}$. Unless $\kappa \gg 1$, this mean free path at $E_{\rm p, max}$ is always much smaller than $R_{\rm j}$.
The condition $\Gamma_{\rm max,NR}(E_{\rm p, max})R_{\rm j}/v_{\rm sh}>5$ was introduced by \cite{Tony_13_Escape} for the case of SNR and where they considered the size of the  spherical shock (instead of $R_{\rm j}$). 
However, \cite{ZirakashviliII_08} found a similar expression for planar non-relativistic shocks. We also note that the same limit applies when Alfv\'en waves dominate given that the maximum growth rate is almost the same \citep{ZirakashviliII_08}.

For large temperatures or low magnetic fields  (see Eq.~\eqref{Tu_th}), thermal effects are important and therefore $E_{\rm p, max}$ is obtained by equating $5/\Gamma_{\rm max,th} = R_{\rm j}/v_{\rm sh}$. If the jet is not completely ionized, the maximum energy of protons due to escape upstream is reduced by a factor $(\Gamma_{\rm max, inc}/\Gamma_{\rm max,NR})^{1/(s-1)}$.

\subsection{Electrons}
\label{Sec_Eemax}

Previous models consider  Alfv\'en turbulence and Bohm diffusion \cite[see e.g.][]{IRAS_07, Valenti_10, Marco_15, Marco_16}.  In the present study, electrons diffuse in the turbulence self generated by protons. They are not the main drivers of the turbulence and can be considered as test particles. If the non-resonant streaming instability gets into the non-linear phase, and possibly other instability can contribute to generate longer wavelength perturbations, we can expect to have the turbulence coherence length $\ell_{\rm c}$ limited by the Larmor radius of the protons at the maximum energy $E_{\rm p,max}$  in the amplified field $B_{\rm s}$ (see the discussion in e.g. \citealt{Reville09, Caprioli_14b}). 

If at first we assume that the maximum electrons energy $E_{\rm e,max}$ exceeds $E_{\rm p,max}$, then the turbulence experienced by electrons will be in the small-scale turbulence regime and  the diffusion coefficient would 
be  $\kappa_{\rm u} \simeq \kappa D_{\rm Bohm}(E_{p,\rm max},B_{\rm s})(E_e/E_{p,\rm max})^2$. This rapid increase with the energy would  limit $E_{\rm e,max}$ to $E_{\rm p,max}$ unless radiative losses dominate and limit it to smaller values. Then, one should expect to have $E_{\rm e,max} \le E_{\rm p,max}$.

By inserting $\kappa_{\rm u}$
into the acceleration timescale ($t_{\rm acc} \propto \kappa_{\rm u}/v_{\rm sh}^2$) and the diffusive loss timescale ($R_{\rm j}^2/6\kappa_{\rm u}$) we find a maximum electron energy $E_{\rm e,max,diff}$  to be within a factor of a few of the maximum electron energy fixed by synchrotron losses $E_{\rm e,max,rad}$ for the different sources in our sample. Then, one can argue that the  statement $E_{\rm e,max} \le E_{\rm p,max}$ is reasonably verified accounting for the uncertainties on the parameters controlled by the microphysics of turbulence generation at fast shocks, namely $\ell_{\rm c}$, $\kappa$, $B_{\rm s}$, and on the macroscopic jet parameters $v_{\rm sh}$ and $R_{\rm j}$. Hereafter we assume that $E_{\rm e,max} = E_{\rm p,max}$. ($E_{\rm e,max,rad} < E_{\rm p,max}$ never happens in our source sample with the derived values of the magnetic field strength $B_{\rm s}$.) We left to a future work more precise calculation of $E_{\rm e,max}$.

\section{Gamma-ray emission}
\label{Sec_gamma}

TeV electrons and protons can emit gamma-rays by their interaction with ambient cold protons through relativistic Bremsstrahlung and proton-proton (pp) collisions. Inverse Compton scattering is a mechanism contributing to gamma-ray emission as well. In the jet termination region, at $\sim 10^{4}$~AU from the central protostar, the stellar photon field is very diluted  and therefore the Inverse Compton scattering is not expected to be very relevant. However, we stress that thermal photons from the bow-shock downstream region can be boosted to the $\gamma$-ray domain. The $\gamma$-ray flux by up-scattering of photons from the bow shock will be computed in a following paper, together with the diffuse emission in the molecular cloud.

Proton-proton and relativistic Bremsstrahlung cooling timescales are comparable. However, 
the number of relativistic protons per unit energy is 
larger than that of electrons, i.e. $N_p > N_e$,  given that  $K_p> K_e$ (see Appendix~\ref{app-2} and Fig.~\ref{Fig_zeta}) and we assume that
the slope is the same in both electrons and protons energy distributions. 
In particular, the distribution of relativistic protons in the jet hotspot is $N_{\rm p} = K_{\rm p} E_{\rm p}^{-s}$, where $K_{\rm p} = K_{\rm e} (m_{\rm p}/m_{\rm e})^{(s-1)/2}$ and  $K_e$ is computed from Eq.~(\ref{Ke_ap}) by fixing $B_{\rm s}=3.3B_{\rm sat,NR}$. In Table~\ref{T:tab1} we list the values of $K_e$ and $K_p$. As a consequence, the emission in the $\gamma$-ray domain will be dominated by hadrons.

\subsection{Proton-proton collisions}

\begin{figure}
\includegraphics[width=\columnwidth]{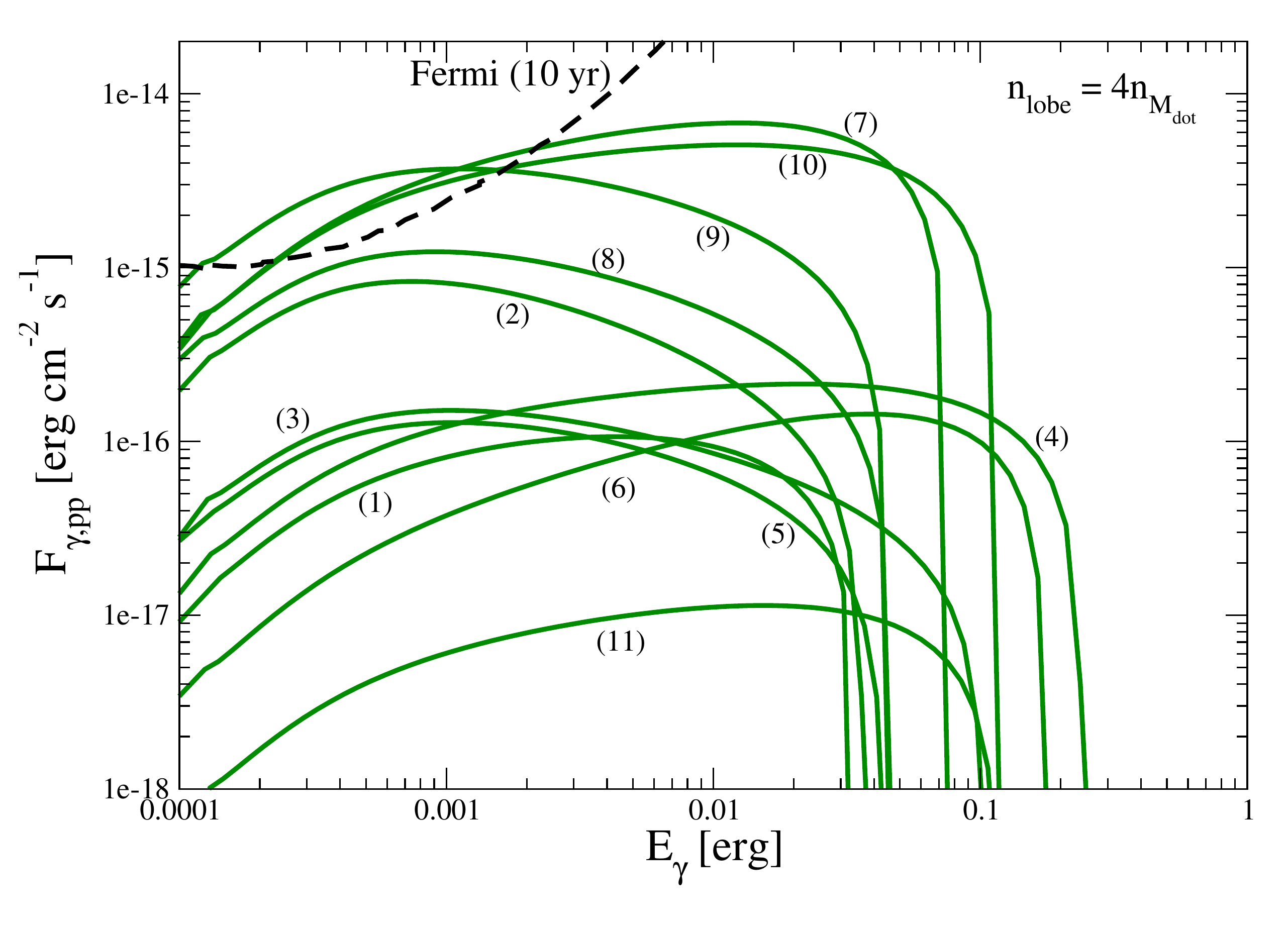}
\caption{$\pi^0$-decay for all the sources in our sample for the case of $n_i=n_{\dot M}$. Black dashed lines indicates the Fermi sensitivity for 10~yr of observation. Numbers indicate the source name (see Table~\ref{T:tab1}).}
\label{sed}
\end{figure}

\begin{figure}
\includegraphics[width=\columnwidth]{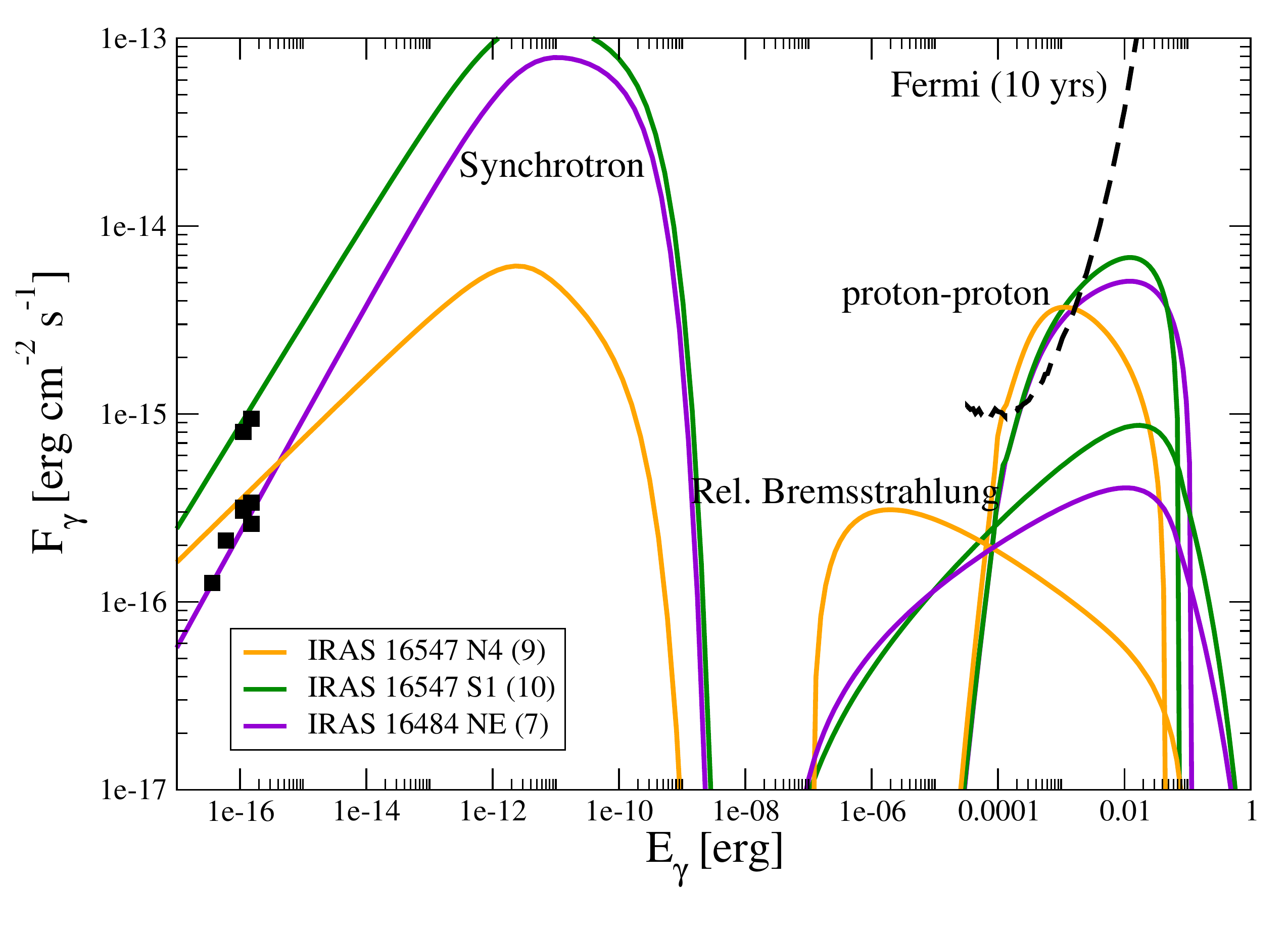}
\caption{ Spectral energy distribution 
from radio to gamma-rays for the three hot spots: IRAS 16547 N4, IRAS 16547 S1, and IRAS 16848 NE. Black squares indicate synchrotron data \citep{Purser_16} and the black dashed line represent the Fermi sensitivity for 10 years of observation.}
\label{sed_total}
\end{figure}

In the one-zone model approximation, the specific luminosity
of gamma rays due to $\pi^0$-decay
can be written as
\begin{equation}\label{L_pp}
E_{\gamma} L_{\gamma, pp} =     E_{\gamma}^2 \,q_{\gamma}(E_{\gamma})\, n_{\rm lobe}
V_e,
\end{equation}
where $q_{\gamma}$ is the emissivity (see Eqs.~(16) and (17) in \cite{IRAS_07}).   We assume that the volume of the $\gamma$-ray emitter is the same as the  synchrotron emission  volume ($V_{\rm e}$) and that the thermal ion density in the lobe is $n_{\rm lobe}=4n_{\dot M}$.
 In Fig.~\ref{sed} we plot the flux $F_{\gamma, \rm pp}=E_{\gamma}L_{\gamma, \rm pp}/(4\pi d^2)$ for all the sources from our sample and the \emph{Fermi}-LAT sensitivity. 
We can see that the predicted emission in G339-8838~NE (7), G34301261~N4 (9), and
S1 (10)  is  detectable by \emph{Fermi}-LAT after 10 years of observation 
(with $5\sigma$-confidence) when $n_{\rm lobe}= 4 n_{\dot M}$.   For  these three sources we plot in Fig.~\ref{sed_total} the
 spectral energy distribution (SED)  including also synchrotron emission and relativistic Bremsstrahlung and using the  formulation in \cite{Blumenthal_Gould_70}. We assume $E_{e,\rm max}=E_{p,\rm max}$ as it is calculated in Eq.~(\ref{Ep-max}).

We note  that $F_{\gamma, pp}\propto n_{\rm lobe}$ and therefore the interaction of relativistic protons with clumps denser than the jet will increase the $\gamma$-ray flux.

\subsection{Density enhancement in the jet termination region}
The termination region of non-relativistic light jets ($n_{\rm j} < n_{\rm mc}$) is expected to be a combination of an adiabatic reverse shock and a radiative bow shock, leading to a large density ratio at the contact discontinuity between both shocks \cite[see e.g.][]{Adriana_Xrays}. The density of the plasma  downstream of the radiative bow shock is 
\begin{equation}\label{n_rad}
\frac{n_{\rm mc}^{\prime}}{n_{\rm mc}} = 500
\left(\frac{v_{\rm bs}}{100\,\rm km\,s^{-1}}\right)^2
\left(\frac{T}{10^4\rm K}\right)^{-1}
\end{equation}
\cite[e.g.][]{Blondin_90} when the plasma is 
cooled down to a temperature $T$, making the density contrast at the contact discontinuity $\chi^{\prime} = 4 n_{\rm j}/n_{\rm mc}^{\prime} \ll \chi$. As a consequence, the contact discontinuity is unstable to dynamical and thermal instabilities. A dense layer of density $n_{\rm mc}^{\prime}$ located at distance $l_{\rm th}$ downstream of the bow shock (see Eq.~\ref{l_th}) fragments into several clumps \cite[e.g.][]{Calderon_20}. In this case, the effective density downstream the reverse shock will increase up to a value $\sim f_{\rm v, clump} n_{\rm mc}^{\prime}$, where $f_{\rm v, clump}\le 1$ is the volume filling factor of clumps in the lobe with volume $\sim R_{\rm j}^3$. 
However, we draw attention to the fact that the component of the magnetic field in the ambient molecular cloud parallel to the bow shock front ($B_{\rm mc,\perp}$) limits the compression factor to a maximum value \citep{Blondin_90}
\begin{equation}\label{B_mc}
\frac{n_{\rm max}^{\prime}}{n_{\rm mc}} \sim 100
\left(\frac{n_{\rm mc}}{10^4\,\rm cm^{-3}}\right)^{\frac{1}{2}}
\left(\frac{v_{\rm bs}}{100\,\rm km\,s^{-1}}\right)
\left(\frac{B_{\rm mc,\perp}}{0.1\rm mG}\right)^{-1} \ .
\end{equation}

The magnetic field in molecular clouds is $B_{\rm mc}\sim$~mG \citep{Crutcher_12} and we have assumed that $B_{\rm mc,\perp}\sim 0.1 B_{\rm mc}$. In the case of light jets, $v_{\rm bs} \sim v_{\rm j}\sqrt{\chi}=v_{\rm j}\sqrt{n_{\rm j}/n_{\rm mc}}$ 
(see Sect.~\ref{Sec_shocks-dens}) and therefore
Eq.~(\ref{B_mc}) can be written as
\begin{equation}\label{B2_mc}
\frac{n_{\rm max}^{\prime}}{n_{\rm mc}} \sim 1000
\left(\frac{n_{\rm j}}{10^4\,\rm cm^{-3}}\right)^{\frac{1}{2}}
\left(\frac{v_{\rm j}}{1000\,\rm km\,s^{-1}}\right)
\left(\frac{B_{\rm mc,\perp}}{0.1\rm mG}\right)^{-1}.
\end{equation}
This indicates that significant enhancement in the plasma density downstream of the reverse shock is feasible if  instabilities grow fast enough to fragment the dense shell and form the clumps. However, we note that even when clumps are not formed, protons accelerated at the jet reverse shock can diffuse down to the dense shell and radiate there.

\subsubsection{Kelvin-Helmholtz and Rayleigh-Taylor instabilities}

Kelvin-Helmoltz (KH) and Rayleigh-Taylor (RT)  instabilities can grow in the contact discontinuity due to the velocity shear and the force exerted by the downstream material of the reverse shock on that of the bow shock. If the bow shock is radiative, the formation of a shell much denser than the jet ($n_{\rm j}$) and the molecular cloud ($n_{\rm mc}$) at the contact discontinuity makes the working surface unstable even in the case of light jets. Following the analysis in \cite{Blondin_90}, the acceleration of the dense shell with a width $w_{\rm shell}$ can be written as 
$a\sim  n_{\rm j} v_{\rm rs}^2 /n^{\prime}_{\rm mc}w_{\rm shell}$. For a
characteristic dynamical timescale $t_{\rm dyn} = R_{\rm j}/v_{\rm j}$ we find
\begin{equation}\label{tau_RT}
\frac{t_{\rm RT}}{t_{\rm dyn}}\sim 0.05
\left(\frac{T}{10^4\rm K}\right)^{-\frac{1}{2}}
\left(\frac{v_{\rm j}}{1000\,\rm km\,s^{-1}}\right)\,,
\end{equation}
where $t_{\rm RT}\sim 1/\sqrt{a k}$ is the growing time of RT instabilities and we have assumed  that $k\sim 1/2w_{\rm shell}$ and $w_{\rm shell}\sim R_{\rm j}/3$.

\section{Summary and conclusions}
\label{sec-conc}

The detection of synchrotron emission from jets powered by high-mass protostars indicates that electrons are in-situ accelerated. We consider a sample of 11
non-thermal lobes in MYSOs as referenced by \citet{Purser_16} and \citet{Obonyo_19}.
From the observational data in the above mentioned papers, the  lobes selected for our study have an average radio spectral index $\alpha = 0.55$, emissivity $\epsilon_{\rm synchr}=2.1\times10^{-30}$~erg~cm$^{-3}$~s$^{-1}$~Hz$^{-1}$, and linear size $R=2.9\times10^{16}$~cm (see Table~\ref{Tab_summary}). Magnetic fields $B_{\rm s} \sim 0.3-5$~mG are needed to explain the synchrotron radio flux. These large values of $B_{\rm s}$ are difficult to explain by a simple compression in an adiabatic strong shock at the jet termination region, at $\sim 0.1$~pc from the central protostar. We note that 
strong compression of the magnetic field in the molecular cloud by the 
radiative bow shock could explain values of $B_{\rm s}$ in the range $0.3-5$~mG , and this scenario will be analysed in a future study. In the present paper we focus on magnetic field amplification by streaming instabilities.

\begin{table}
\caption{Average values of the most relevant parameters in our study (see Tables~\ref{Tab_dens} and \ref{T:tab1}).}\label{Tab_summary}
\begin{tabular}{ll}
\hline\hline
\multicolumn{2}{c}{From observations}\\
\hline
Radio spectral index & $\alpha=0.55$\\
Synchrotron emissivity  & $\epsilon_{\rm syn, \nu}=2\times10^{-30}$~erg~cm$^{-3}$ s$^{-1}$~Hz$^{-1}$ \\
Size of the non-thermal lobe  & $R=2.9\times10^{16}$~cm \\
\hline
\multicolumn{2}{c}{Assumed values}\\
\hline
Shock speed   &$v_{\rm sh}=1000$~km~s$^{-1}$ \\
Jet magnetic field & $B_{\rm j}=10\,\mu$G \\
Jet temperature & $T_{\rm u}=10^4$~K \\
\hline
\multicolumn{2}{c}{Computed values}\\
\hline
\multirow{2}{*}{Jet ion density} & $n_{\rm i}=n_{\dot M} = 1.2\times10^3$~cm$^{-3}$\\
& $n_{\rm i}=\langle n_{i} \rangle = 1.6\times10^4$~cm$^{-3}$\\
Minimum shock speed   &$v_{\rm sh,ad}=455$~km~s$^{-1}$ \\
Equipartition magnetic field & $B_{\rm eq}=7.25$~mG\\
Downstream magnetic field & $B_{\rm s}=0.84$~mG\\
Protons total energy density & $U_{p,\rm tot}=1.5\times10^{-5}$~erg~cm$^{-3}$  \\
Protons total acceleration efficiency & $\eta_{p,\rm tot}=0.05$  \\
Protons maximum energy & $E_{p,\rm max}=0.24$~TeV\\
 Electrons normalization constant & $K_e = 3.8\times10^{-9}$~erg$^{s-1}$~cm$^{-3}$ \\
Protons normalization constant & $K_p = 2.4\times10^{-7}$~erg$^{s-1}$~cm$^{-3}$ \\
\hline
\hline
\end{tabular}
\end{table}

CR streaming instabilities can generate magnetic field perturbations necessary for the particles to diffuse back and forth the shock. The CR streaming can amplify resonant Alfv\'en waves with wavelength of the order of the CR Larmor radius $r_{\rm g}$ but also excite the NR (Bell) instability which produces perturbations at scales much smaller than $r_{\rm g}$  \citep{Tony_04}. Bell's instability can amplify small perturbations in the magnetic field up to values much larger than the unperturbed jet magnetic field $B_{\rm j}$ if
$\zeta M_A > 1$ where
\begin{equation}\label{Eq:driving}
\zeta M_{\rm A}^2 = 7000
\left(\frac{U_{\rm p}}{10^{-7}\rm erg\,cm^{-3}}\right) 
\left(\frac{B_{\rm j}}{10 \mu\rm G}\right)^{-2}
\left(\frac{v_{\rm sh}}{1000\, \rm km\,s^{-1}}\right)
\end{equation}
is an equivalent expression for Eq.~(\ref{Eq:driving_tot}).
We remark that even in the case where $\zeta$ is small due to the slow speed of the shock (compared e.g. with supernova remnants), the large values of jet densities makes $M_{\rm A}$ significantly large to satisfy the condition $\zeta M_{\rm A}^2 \gg 1$ over a large range of parameters. However, large ion densities in a non-completely ionized jet would increase the ion-neutral collisions. Alfv\'en waves can be damped by ion-neutral collisions  \cite[e.g.][]{Drury_damping}. Conversely, Bell modes are not as heavily damped, although the maximum growth rate decreases \citep{Brian_07}. Thermal effects only weakly affect the NR instability growth rate (see Fig.~\ref{growth}). High magnetic field obliquity can also lead to a decrease of the streaming instability growth rates or even a complete quenching if the precursor length becomes too short (see Sect.~\ref{S:Obl}). 

By assuming that the large magnetic field in the  synchrotron emitter ($B_{\rm s}$) is due to amplification through the Bell's instability and fixing $v_{\rm sh}=1000$~km~s$^{-1}$ we estimate $B_{\rm s}\sim 0.4B_{\rm eq}$ and the energy density in non-thermal electrons $U_{e,\rm tot}$. Then, the energy density in non-thermal protons is $U_{p,\rm tot} = a U_{e,\rm tot}$. Under the assumption that the jet ionized density is
$n_i = \langle n_i \rangle$, we estimate the proton acceleration efficiency $\eta_{p, \rm tot}$ to be $\sim 0.05$. We stress that this method is different with respect to that used in supernova remnants, where the magnetic field is usually estimated by comparing the width of X-ray filament profiles with the synchrotron cooling length.

By knowing $\eta_{p, \rm tot}$ and $B_{\rm s}$ we estimate the 
maximum energy of protons accelerated in the jet reverse shock and the $\gamma$-ray emission that they produce.
By considering the amplification timescale of the magnetic field in the shock upstream region we find $E_{p,\rm max}\sim 0.1$~TeV. These protons can emit $\gamma$ rays through their collisions with thermal ions. 
We compute the  $\gamma$-ray flux $F_{\gamma, pp}$ for all the sources in our sample.
We note that $F_{\gamma, pp}\propto n_{\rm lobe}$ and the jet ionized density at the termination region is $n_{\dot M}\le n_i \le n_{\rm ff}$, where $n_{\rm ff}/n_{\dot M} \sim 100$ (see Fig.~\ref{vshock-min}).
We find that having a density
$n_{\rm lobe} = 4 n_{\dot M}$ in the non-thermal emitter 
is  high enough to reach detectable levels of $\gamma$ rays  with \emph{Fermi} in IRAS 16547 N4, IRAS 16547 S1, and IRAS 16848 NE, as we show in Fig.~\ref{sed}. 
Although there is no claim of detection of these sources by Fermi, we expect that our result will motivate a future study on the Fermi data at the location of these sources. In addition, these sources will be perfect targets for a point-source mode. We also note that 
mixing due to dynamical instabilities can significantly enhance the density of targets in the lobes. The very dense shell formed downstream of the radiative bow shock is unstable and fragmented in clumps with density $\sim 100-1000$ times the density of the ambient medium (i.e. the molecular cloud).
By achieving $F_{\gamma, pp}\sim 5\times10^{-13}$~erg~cm$^{-2}$~s$^{-1}$ at 0.1~TeV,
CTA will be of great importance to measure the cut-off of the  spectrum.
Also, the spatial resolution of CTA is expected to be better than Fermi. The detection of $\gamma$ rays from protostellar jets will open a new window to study stellar formation, as well as the efficiency of  DSA in the high density ($n_{\rm i}\sim 10^3-10^4$~cm$^{-3}$) and low velocity ($v_{\rm sh}\sim 1000$~km~s$^{-1}$) regime.  In particular, the detection of diffuse gamma-ray emission in molecular clouds where MYSOs are embedded will be a piece of evidence of proton acceleration in protostellar jets.

In a following paper we will perform more detailed calculations of the multi-wavelength lepto-hadronic spectral energy distribution, from radio to $\gamma$ rays. The diffuse $\gamma$-ray emission of particles accelerated in the jet termination shocks and interacting with ions in the molecular cloud where the protostar is embeded will be also modeled, as well as the emission of secondary particles. We will select the most promising candidates from the sample of sources in the present study, as well as other sources such as IRAS 18162-2048, the powering massive protostar of the Herbig Haro objects HH80, HH81 and HH80N \cite[see e.g.][]{Adriana_17,Adriana_Xrays}.

High sensitivity radio observations in the GHz domain are very important to model the synchrotron emission and constrain the energy density in non-thermal particles (Sect.~\ref{S:Rel}).  To this purpose, the detection of polarized emission will be crucial to disentangle thermal contamination in the GHz domain, as well as to investigate the morphology of the magnetic field near the shock. The Next Generation Very Large Array (ngVLA) will play a fundamental role on that \citep{Galvan_Madrid_18,Hull_18}.
Moving to lower frequencies, the low-energy cutoff of the electrons distribution is an important parameter related with the efficiency of shocks to inject electrons from the thermal pool to the high energy tail. In this sense, the forthcoming Square Kilometre Array (SKA) will be extremely important to observe the low-energy cutoff \citep[see e.g.][]{cut-off}.

Finally we note that protostellar jets have been well studied through several laboratory experiments \cite[e.g.][]{Nicolai_08,Liang_18,Suzuki-Vidal_12}. 
In particular, \cite{Suzuki-Vidal_15} have shown that laboratory bow shocks formed by the collision of two counterstreaming and supersonic plasma jets is fragmented due to the rapid growth of thermal instabilities. The formation of collisionless shocks \citep{PhysRevLett.123.055002}, the development of plasma instabilities, and the acceleration of particles in laser plasma \citep{Reville_13} is going to open a fascinating era of laboratory astrophysics  in synergy with high energy-astrophysics.

\section*{Acknowledgements}
The authors thank the anonymous referee for his/her comments  that helped us to improve the quality of our paper.  
The authors thank C. Carrasco-Gonz\'alez and S. Cabrit for insightful discussions on protostellar jets.  
A.T.A. thanks Tony Bell and K.~Blundell for encouragement and motivating discussions on jets, particle acceleration, and streaming instabilities.  A.T.A. thanks
M.V. del Valle and R. Santos-Lima for their help with numerical calculations. 
A.T.A. thanks the Czech Science Foundation under the grant GA\v{C}R 20-19854S titled ``Particle Acceleration Studies in Astrophysical Jets''.
M.P. acknowledges funding from the INAF PRIN-SKA 2017 program 1.05.01.88.04,
by the Italian Ministero dell'Istruzione, Universit\`a e Ricerca through the grant Progetti Premiali 2012--iALMA (CUP C52I13000140001).
and by the project PRIN-INAF-MAIN-STREAM 2017 ``Protoplanetary disks seen through the eyes of new-generation instruments''.
This work has been carried out thanks to the support of the
OCEVU Labex (ANR-11-LABX-0060) and the A*MIDEX project (ANR-11-
IDEX-0001-02) funded by the "Investissements d'Avenir" French government
program managed by the ANR.




\bibliographystyle{mnras}
\bibliography{yso.bib} 



\appendix

\section{Synchrotron emission}
\label{app-1}

\begin{figure}
\includegraphics[width=\columnwidth]{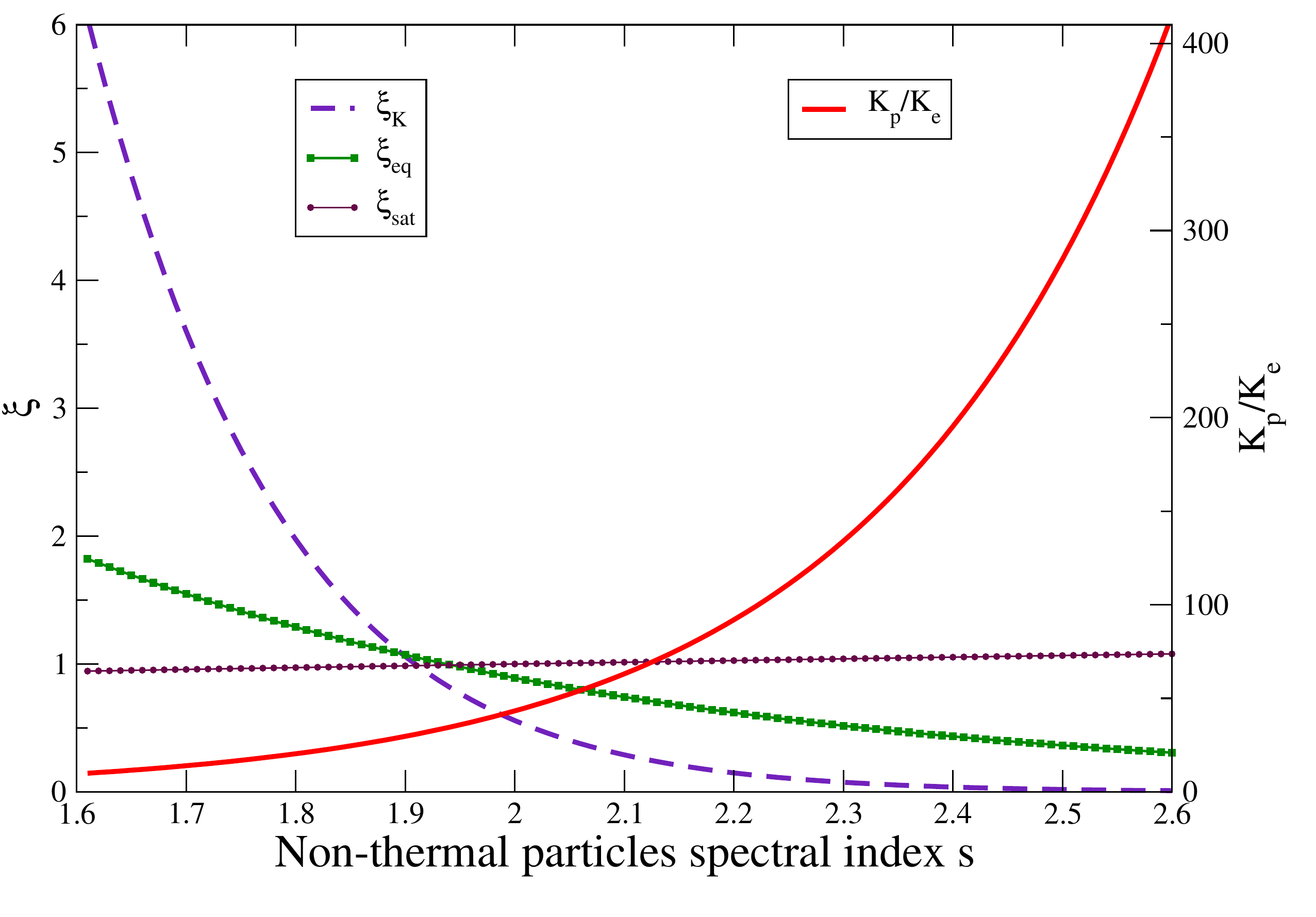}
\caption{Functions $\xi_K$ (Eq.~\ref{g_s}), $\xi_{\rm eq}$, and $\xi_{\rm sat}$ (left y-axis) and $K_p/K_e$ (right y-axis)
as a function of the spectral index of non-thermal particles.}
\label{Fig_zeta}
\end{figure}

The synchrotron emissivity per unit frequency $\nu$ emitted by
a source with volume $V_{\rm e}$ located at distance $d$ from Earth is $\epsilon_{\rm syn,\nu} = 4\pi d^2 S_{\nu}/V_{\rm e}$, giving
\begin{equation}\label{emiss_nu}
\frac{\epsilon_{\rm syn,\nu}}{\rm erg \,Hz^{-1} s^{-1} cm^{-3}}= 
\frac{1.2\times 10^{-30}}{f_{\rm Ve}}
\left(\frac{d}{\rm kpc}\right)^2
\left(\frac{S_{\nu}}{\rm mJy}\right)
\left(\frac{R}{10^{16} \rm cm}\right)^{-3},
\end{equation}
where $S_{\nu}$ is the flux at frequency $\nu$. We have defined $V_e = f_{\rm Ve} R_{\rm j}^3$, where $V_e$ is the volume filled in with non-thermal electrons and $f_{\rm Ve}$ is the volume filling factor. We consider that $\epsilon_{\rm syn,\nu} = 4\pi \varepsilon_{\nu}$, where the emission coefficient for synchrotron radiation is
\begin{equation}
    \varepsilon_{\nu} = c_5 K_e \langle \sin \Theta^{\frac{s+1}{2}}\rangle B_s^{\frac{s+1}{2}}\left(\frac{\nu}{2 c_1}\right)^{\frac{1-s}{2}},
\end{equation}
$\Theta$ is the electron pitch angle,  $c_1 = 6.264\times10^{18}$ Hz  
and 
\begin{equation}\label{c5}
c_5 = {1.866\times10^{-23} \over (s+1)}\Gamma\left(\frac{3s-1}{12}\right)\Gamma\left(\frac{3s+19}{12}\right)\,{\rm erg\,G^{-1}\,sterad^{-1}} 
\end{equation}
\citep{Arbutina_12,revised_equip}. We assume an isotropic distribution for the orientation
of $\Theta$ and therefore
\begin{equation}\label{pitch}
    \langle \sin \Theta^{\frac{s+1}{2}}\rangle = \frac{\sqrt{\pi}}{2}\frac{\Gamma\left(\frac{s+5}{4}\right)}{\Gamma\left(\frac{s+7}{4}\right)}.
\end{equation}
By combining Eqs.~(\ref{emiss_nu})-(\ref{pitch}) we find that 
\begin{eqnarray}\label{Ke_ap}
\frac{K_{e}}{\rm erg^{s-1} \,cm^{-3}} & \approx &1.6\times10^{-9} \frac{\xi_K(s)}{f_{\rm Ve}}
\left(\frac{d}{\rm kpc}\right)^2
\left(\frac{S_{\nu}}{\rm mJy}\right)
\left(\frac{R}{10^{16}\rm cm}\right)^{-3}\nonumber\\
&&
\times \left(\frac{\nu}{\rm GHz}\right)^{\frac{s-1}{2}}
\left(\frac{B_{\rm s}}{\rm mG}\right)^{-\frac{s+1}{2}},
\end{eqnarray}
where
\begin{equation}\label{g_s}
\xi_K(s) \simeq  10^{-3.55(s-2)}  \frac{\Gamma\left(\frac{s+5}{4}\right)\Gamma\left(\frac{3s-1}{12}\right)\Gamma\left(\frac{3s+19}{12}\right)}{\Gamma\left(\frac{s+7}{4}\right) (s+1)}.
\end{equation}
The function $\xi_K(s)$ is plotted in Figure~\ref{Fig_zeta}.

\section{Particle energy distribution in the shock downstream region}
\label{app-2}

\begin{figure}
\includegraphics[width=\columnwidth]{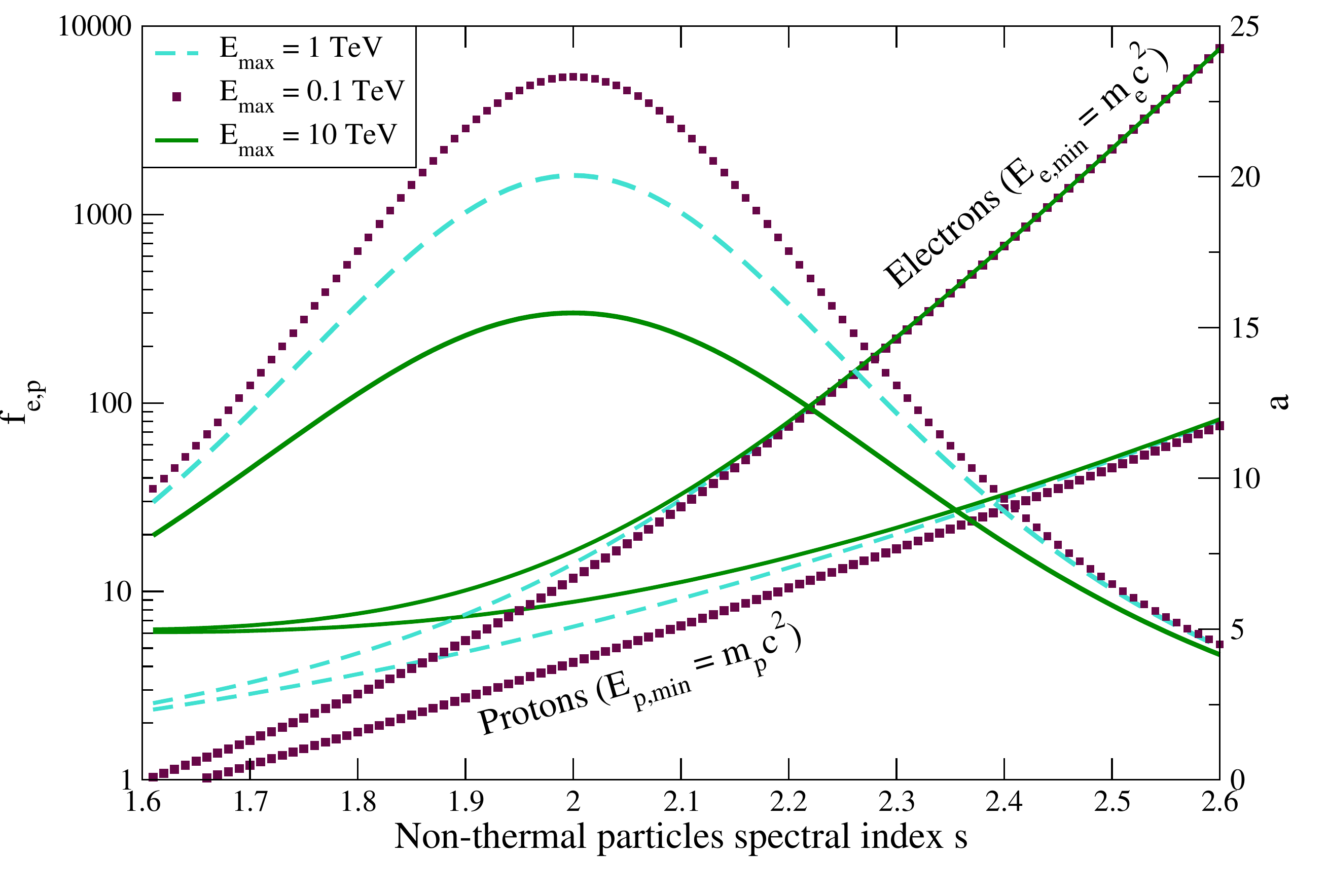}
\caption{Functions $f_e$ and $f_p$ (left axis) and $a$ (right axis)  as a function of the spectral index of non-thermal particles
for different values of 
$E_{e,\rm max}$ and $E_{p,\rm max}$.}
\label{Fig_fs}
\end{figure}

Non-thermal particles accelerated in a non-relativistic shock follow a power-law energy distribution
\begin{equation}\label{spectrum}
N_{\rm e,p} = \left\{ \begin{array}{ll}
K_{\rm e,p}{m_{\rm e,p}c^{2}}^{\frac{(1-s)}{2}}E_{\rm e,p}^{-\frac{s+1}{2}} & E_{\rm inj}<E_{\rm e,p} < m_{\rm e,p}c^2,\\
 K_{\rm e,p} E_{\rm e,p}^{-s}&  m_{\rm e,p}c^2 <E_{\rm e,p} <E_{\rm e,p,\rm max},
\end{array}\right.  
\end{equation}
where $e$ and $p$  stands for electrons and protons, respectively, and
$E_{\rm inj}\sim 2 m_pv_{\rm sh}^2$.
We note that the condition $N_e=N_p$ at $E < m_e c^2$ gives $K_p/K_e = (m_p/m_e)^{(s-1)/2}$ (see Fig.~\ref{Fig_zeta}).
The spectrum is flatter in the non-relativistic regime, and therefore non-relativistic particles do not contribute significantly to  the total energy density $U_{\rm e,p}$ and hence
most of the  non-thermal number and energy density is due to particles with 
 $E_{\rm e,p} \gtrsim m_{\rm e,p}c^2$. 
In the case of negligible energy losses, the total energy density stored in non-thermal particles is $U_{e,p,\rm tot} = K_{e,p} f_{e,p}(s)$, where
\begin{equation}\label{f_gamma}
f_{e,p} \sim \left\{ \begin{array}{ll}
\left(\frac{1}{2-s}\right) E_{e,p,\rm max}^{2-s} & \text{if } s<2\\
\log\left(\frac{E_{e,p,\rm max}}{m_{e,p} c^2}\right) & \text{if } s=2 \\
\left(\frac{1}{s-2}\right) (m_{e,p} c^2)^{2-s} & \text{if } s>2.
\end{array}\right.  
\end{equation}
We note that the ratio $a=U_{p,\rm tot}/U_{e\rm tot}$ can be written as $a=(m_p/m_e)^{(s-1)/2} f_p/f_e$, and for  $s>2$ it can be approximated as $a\sim(m_p/m_e)^{(3-s)/2}$. In Figure~\ref{Fig_fs} we plot $f_p$, $f_e$, and
$a$ for $1.6 \le s\le 2.6$ and
$E_{e,\rm max} =E_{p,\rm max} = 0.1$, 1, and 10~TeV.

The energy and number density of particles with a certain energy $E_{\rm e,p}$ are $n_{e,p} = K_{e,p} E_{e,p}^{1-s}$ and $U_{e,p} = K_{e,p} E_{e,p}^{2-s}$, respectively. We are particularly interested in protons, 
given that they are responsible for the CR streaming instability. The mumber and energy  density of protons with energy $E_{p}$ are
\begin{equation}\label{n_gamma}
n_{p} =\frac{U_{p,\rm tot}}{f_p}E_p^{1-s} \sim \left\{ \begin{array}{ll}
\left(\frac{U_{p,\rm tot}}{E_{p,\rm max}}\right)(2-s) \left(\frac{E_p}{E_{p,\rm max}}\right)^{1-s} & \text{if } s<2\\
\frac{U_{p,\rm tot}}{m_pc^2 \log\left(\frac{E_{p,\rm max}}{mc^2}\right)} \left(\frac{E_p}{m_pc^2}\right)^{-1} & \text{if } s=2 \\
\left(\frac{U_{p,\rm tot}}{m_p c^2}\right) (s-2) \left(\frac{E_p}{m_pc^2}\right)^{1-s} & \text{if } s>2
\end{array}\right.  
\end{equation}
and
\begin{equation}\label{U_gamma}
U_{p} =\frac{U_{p,\rm tot}}{f_p}E_p^{2-s} \sim \left\{ \begin{array}{ll}
U_{p,\rm tot}\,(2-s) \left(\frac{E_p}{E_{p,\rm max}}\right)^{2-s} & \text{if } s<2\\
\frac{U_{p,\rm tot}}{\log\left(\frac{E_{p,\rm max}}{m_p c^2}\right)} & \text{if } s=2 \\
U_{p,\rm tot}\, (s-2) \left(\frac{E_p}{m_pc^2}\right)^{2-s} & \text{if } s>2,
\end{array}\right. 
\end{equation}
respectively. 


\bsp	
\label{lastpage}
\end{document}